\documentclass[a4paper,11pt]{article}
\usepackage{jheppub} % for details on the use of the package, please see the JINST-author-manual
\usepackage{lineno}
\usepackage{amsmath}
\usepackage{amsfonts}
\usepackage{graphicx}
\usepackage{cancel}
\usepackage{stmaryrd}
\usepackage{graphicx}
\usepackage{float}
\usepackage{caption}
\usepackage{subcaption}
\usepackage{feynman}
\usepackage{pifont}
\title{Quantitative classicality in cosmological interactions during inflation \\}

\author[a]{Yoann L. Launay,}

 \affiliation[a]{Centre for Theoretical Cosmology, Department of Applied Mathematics and Theoretical Physics,
University of Cambridge, Wilberforce Road, Cambridge CB3 0WA, United Kingdom}
\emailAdd{yoann.launay@outlook.com} 

\author[b]{Gerasimos I. Rigopoulos,}
\affiliation[b]{School of Mathematics, Statistics and Physics,  Newcastle University, Newcastle upon Tyne, NE1 7RU, United Kingdom}
\emailAdd{gerasimos.rigopoulos@newcastle.ac.uk}

\author[a]{E. Paul S. Shellard}
\emailAdd{eps1@cam.ac.uk}
\abstract{
We examine the classical and quantum evolution of inflationary cosmological perturbations from quantum initial conditions, using the on-shell and off-shell contributions to correlators to investigate the signatures of interactions. In particular, we calculate the Keldysh contributions to the leading order bispectrum from past infinity, showing that the squeezed limit is dominated by the on-shell evolution. By truncating the time integrals in the analytic expressions for contributions to the bispectrum, we define a `quantum interactivity' and  quantitatively identify scales and times for which it is sufficient to only assume classical evolution, given a fixed precision. In contrast to typical perceptions inspired by free two-point functions, we show that common non-linear contributions to inflationary perturbations can be well-described by classical evolution even prior to horizon crossing. The insights gained here can pave the way for quantitative criteria for justifying the validity of numerically simulating the generation and evolution of quantum fluctuations in inflation. In particular, we comment on the validity of using stochastic inflation to reproduce known in-in perturbative results. An extensive appendix provides a review of the Keldysh formulation of the in-in formalism with the initial state set at a finite, as opposed to infinite past, emphasizing the importance of considering temporal boundary terms and the initial state for correctly obtaining the propagators. We also show how stochastic dynamics can emerge as a sufficiently accurate approximation to the full quantum evolution. This becomes particularly transparent in the Keldysh description. 
}

\begin{document}
\maketitle
\flushbottom

\section{Introduction \label{sec:intro}}

The theory of inflation \cite{starobinsky80,guth_1981,Linde82} is accepted as a natural mechanism to create density inhomogeneities \cite{Mukhanov1981,Hawking1982,Starobinsky1982pert,Guth82pert,HawkingMoss1982,Bardeen83} 
as primordial seeds for cosmic structures observed today while solving the issues of the old Big Bang. Its final two-point statistics for the density or curvature perturbations are also compatible with the cosmic microwave background (CMB) data to very high precision \cite{PlanckInflation}. Non-Gaussianity in the higher-order spatial correlation functions of the cosmological fluid constitutes one of the key predictions of inflation via its interactions, including gravity 
\cite{Maldacena_2003}. Different inflationary models produce different non-Gaussian signatures that are already constrained by the CMB \cite{PlanckNG}, but forthcoming galaxy surveys promise measurements which may bring us closer to potential detection. Whether detected or not, these experiments will continue to offer insights about possible physical scenarios describing the early universe. 

The core assumption of inflation lies in the contention that the two-point statistics of the inhomogeneities trace their origin back to the quantum vacuum fluctuations on small scales. Treating these inhomogeneities perturbatively around a quasi-de Sitter spacetime is permitted in general relativity thanks to the use of well-established quantum field theory on curved spacetime (QFTCS) \cite{WaldQFTCS,Weinberg95}. However, the focus on the tools of quantum theory on curved spacetime might have left out a simple question: how necessary is the quantum treatment for describing cosmological observables and thus also for data and simulations? %For instance, it is already known that two-points statistics can be tricked \cite{Martin16}.

{As established previously \cite{Weinberg05,Weinberg06}, quantum contributions are limited in growth and leave a universe behaving classically on scales roughly larger than the Hubble radius. More generally, the concepts of classicality and quantumness can have different measures depending on the context, many of which have been applied to cosmology, from Weinberg's criterion to decoherence \cite{Burgess_2023}, 
 quantum purity \cite{Colas24purity} or quantum discord \cite{Ollivier01}, and many other criteria. However, classical behaviour specified according to one approach may or may not be classical according to another. For instance, a state can show very high entanglement and still have negligible non-commuting observables \cite{Martin16}.
 A lot of these studies have mostly looked at the scalars in a fixed (quasi-) de Sitter background and concentrated on the late-time limit, but it is desirable to explore this behaviour more quantitatively and to obtain precise times for the onset of classicality for a given model.}

To understand our choice of what classicality means in this study, it is important to note a growing number of ever more ambitious numerical studies of inhomogeneous inflation such as lattice simulations \cite{latticeeasy,Easther10,Cosmolattice,Caravano22lattice,Caravano24butterfly,caravano2024axionbackreaction, caravano2024usr}, numerical relativity \cite{East16,clough_robustness_2017,Bloomfield19,aurrekoetxea_effects_2020,Corman23, elley2024}, and stochastic inflation \cite{Jackson22,cruces_stochastic_2022,tomberg_numerical_2023,Launay24,stolas,Animali25}. These studies constitute a great leap in theory predictions because of the possibility to compute correlation functions for any models of inflation and at all orders in perturbation theory. {These numerical systems all evolve according to the classical equations of motion (EOM), sourced by initial or stochastic noise, the spectra of which are motivated by the aforementioned quantum fluctuations. Unfortunately, the precise time from which the non-linear dynamics can be described by the EOM is usually unknown or ignored. This work will show that there can be an answer for each scenario and regime. With these interests in mind, let us emphasize the following convention:
\begin{center}
    In this work, `classical' or `on-shell' refers to any evolution of initial or stochastic sources with the classical EOM.
\end{center}
Strictly speaking, and as we will see, it is thus a form of semi-classicality.}

In what follows, Section \ref{sec:sec1} showcases the Keldysh rules, starting from the Dyson \textit{in-in} formula. The on-shell and off-shell contributions to the bispectrum are computed explicitly and analysed. Section \ref{sec:sec2} focuses on finding scales, regimes and methods which can guarantee the domination of on-shell contributions to the bispectrum. In particular, a few remarks are made about the validity of stochastic inflation. We conclude in Section \ref{sec:Conclusion}. This work relies on the Keldysh formulation of the in-in formalism which we review in Appendix \ref{app:CTP}, paying attention to the inclusion of an initial state set at a finite past. Appendix \ref{app:CSA} reviews the origins of the classical-statistical approximation while Appendix \ref{app:appRedef} recalls contributions to the bispectrum arising from field redefinitions.   

\paragraph{Conventions.} 

$M_{Pl}$ is the \textit{reduced} Planck mass.

$t$ is the cosmological time, as opposed to $\eta$ the conformal time.

$\partial$ designates a comoving coordinate derivative (no $a$ scaling).

$\hat{\cdot}$ designates an operator in the QFT sense and its dependence determines its picture.

$\tilde{\cdot}$ designates a random variable, e.g. a stochastic source.

$\langle\cdot\rangle'$ indicates an $n$-point function in Fourier space with an implicit $\delta^{(3)}(\boldsymbol{k_1}+...+\boldsymbol{k_n})$ factor. The 3D space of positive momenta such that $\boldsymbol{k_1}+...+\boldsymbol{k_n} = 0$ (up to some $k_\textrm{max}$) will be referred to as the \textit{tetrapyd}.

Three-dimensional spatial vectors (most likely comoving coordinates) are written as bold symbols.

A Green's function will always be defined as the solution to the standard Green's equation. In particular for two scalar fields $\phi_a$ and $\phi_b$, we will always have in both real and Fourier space $iG^{ab}(u,v) =  \langle\hat{\phi}_a(u)\hat{\phi}_b(v)\rangle$. 

\section{On-shell and off-shell bispectra \label{sec:sec1}} 
\subsection{Cosmological framework}
 In this work and unless specified, we focus on a typical choice of early universe scalar cosmological action. It is that of an inflaton $\phi$ in a $P(X,\phi)$ field theory treated in general relativity
\begin{equation}
    S = \frac{1}{2}M_{Pl}^2\int d^4x \sqrt{-g} R+ \int d^4x \sqrt{-g} P(X,\phi),
\end{equation}
where $X = -\frac{1}{2}\partial^{\mu}\phi\partial_{\mu}\phi$, yielding a speed of sound $c_s = P_{,X}/( P_{,X}+2X P_{,XX})$. This is usually studied in perturbation theory in an appropriate time-slicing (comoving gauge) to give the perturbation action up to cubic order \cite{chen_observational_2007} 
\begin{eqnarray}
S &=& \int d^4x \,  \frac{1}{2}\frac{az^2}{c_s^2}(\dot{{\cal R}}^2-\frac{c_s^2}{a^2}\left (\partial {\cal R})^2 \right ) +S^{(3)} +{\cal O}({\cal R}^4) \nonumber \\
  &\equiv& S^{(2)}+S^{(3)} +{\cal O}({\cal R}^4) 
\label{eq:linearisedAction}
\end{eqnarray}
where ${\cal R}$ is defined such that $g_{ij} = a^2 e^{2{\cal R}}\delta_{ij}$ in comoving gauge, and the background quantity $z^2(t) = a^2H^{-2}\dot{\phi}^2 = 2a^2\varepsilon_1M_{Pl}^2$, where $\varepsilon_1 = -H^{-1}d_t\ln H$ is the background slow-roll parameter. This perturbative truncation will be taken as our dynamical action, where the cubic $S^{(3)}$ contribution will be made explicit later. Note that the discussions laid down in this article are extendable beyond GR, for instance with the EFT of inflation \cite{Cheung2008}.

In this framework, the free theory's equation of motion is ${\cal D}_0{\cal R} = 0$, usually referred to as the Mukhanov-Sasaki equation \cite{Mukhanov1981}, where the operator $\mathcal{D}_0$ is expressed as 
\begin{equation}
\mathcal{D}_0  = -\partial_t\left (\frac{az^2}{c_s^2}\partial_t\cdot\right)+\frac{z^2}{a}\partial^2\,.
\end{equation}
The initial state is often chosen as the Bunch-Davies (BD) vacuum \cite{BD}, described by the wave-functional
\begin{equation}
    \langle{\cal R}^{\rm in}|0\rangle = \mathcal{N}\exp\left(-\frac{1}{2}\int d^3\mathbf{x}d^3\mathbf{y}\, {\cal R}^{\rm in}(\mathbf{x})\frac{\mathcal{E}(\mathbf{x},\mathbf{y})}{\gamma}{\cal R}^{\rm in}(\mathbf{y})\right),
    \label{eq:vacwave}
\end{equation}
where
\begin{equation}
    \frac{\mathcal{E}(t,\boldsymbol{x},\boldsymbol{y})}{\gamma} = \frac{z^2(t)}{\gamma}\int \frac{d^3\boldsymbol{k}}{(2\pi)^3}e^{i\boldsymbol{k}\cdot (\boldsymbol{x}-\boldsymbol{y})}\omega_k,
    \label{eq:BDamp}
\end{equation}
%which has inverse bulk two-point function at past infinity $t_{\rm in} = -\infty$
%as opposed to the conjugate momentum one writing as $(\gamma{\cal E})^{-1}$, 
with $\omega_k^2 = k^2$ and ${\cal R}^{\rm in} = {\cal R}(t_{\rm in})$ for some initial time $t_{\rm in}$, often taken $t_{\rm in} \rightarrow -\infty$ for analytical computations. The parameter $\gamma$ is dimensionless and serves as a book-keeping device to track places where $\hbar$ would appear had we kept SI units. It will serve as a signifier of contributions that cannot be reproduced by on-shell evolution, i.e.~arising from the classical equations of motion. It should be set to unity when not required, $\gamma\rightarrow 1$. The above wave functional is derived in the $k\gg aH $ imit, where the theory is free and massless, by following usual procedures \cite{Weinberg95,Chen17diag, Kaya19}. In that sense, we recover harmonic quantum mechanics for each mode. In this work we will also consider a finite $t_{\rm in}$ sufficiently early such that \eqref{eq:BDamp} is an accurate desciption. 

When looking at the example of a quasi-dS spacetime evolution with such initial conditions (ICs), one can find analytical solutions of the Mukhanov-Sasaki equation as
\begin{equation}
    {\cal R}_k(\eta) = -\frac{H}{M_{Pl}}\sqrt{\frac{\pi \gamma}{8\varepsilon_1 k^3}}(-k \eta)^{3/2}H^{(1)}_{\nu}[-k\eta]
    \label{eq:quasiDShankel}
\end{equation}
which can be approximated as a quasi massless solution in the slow-roll limit ($\varepsilon_{1,2} \ll 1 \Rightarrow \nu\simeq 3/2^+$) to recover the well-known solution
\begin{equation}
    {\cal R}_k(\eta) = \frac{iH}{M_{Pl}}\sqrt{\frac{\gamma}{4\varepsilon_1 k^3}}(1+i k \eta)e^{-ik\eta}
    \label{eq:quasiDS}
\end{equation}
where $c_s$ has been set to $1$ for simplicity. This leads to the usual formula for the free two-point spacetime correlation function $\langle \hat{\cal R}(x)\hat{\cal R}(y)\rangle$.

\subsection{Keldysh \textit{in-in} formula}

QFTCS can be studied using various formalisms: The closed-time-path (CTP) or in-in formalism tracing back to the original works of Keldysh \& Schwinger   \cite{Schwinger60,Bakshi62,Bakshi63,Keldysh64}, the wavefunctional formalism \cite{Maldacena_2003} or more recently bootstrapping \cite{arkani-hamed_cosmological_2020}. All results are consistent for the same scenario but show different advantages depending on the study. In this work, we will use the first formalism, one formulation of which can be summarised by the \textit{in-in} formula for correlators of a quantum field $\hat{\cal R}$ given an initial state, here chosen to be the BD vacuum $| 0 \rangle $  
\begin{equation}
    \langle \hat{\cal R}(x_1)...\hat{\cal R}(x_n) \rangle = \langle 0 |  {\rm \Bar{T}}\exp\bigg(\frac{i}{\gamma}\int_{\cal C}{\cal H}_{int}[\hat{\cal R}_-^{\cal I}]\bigg) \hat{\cal R}^{\cal I}(x_1)...\hat{\cal R}^{\cal I}(x_n) {\rm T}\exp\bigg({-\frac{i}{\gamma}\int_{\cal C} {\cal H}_{int}[\hat{\cal R}_+^{\cal I}]}\bigg) | 0\rangle,
    \label{eq:ininqcl}
\end{equation}
where the integration is over ${\cal C } = [t_{\rm i},t_{\rm f}]\times \mathbf{R}^3$, and ${\rm T}$ (resp. ${\rm\bar{T}}$) denote the time ordering operator (resp. anti-time ordering), acting on the exponentiated contributions of the interactions over time. The evolution operator appears twice, once for the ordered field and once for the anti-ordered ones, reflecting the ket and bra state vectors involved in quantum mechanical expectation values. We have made explicit the two $\pm$ (forwards and backwards temporal evolution) branches of the field, usually kept implicit and encoded in the contours' $\pm i\varepsilon$ prescription \cite{Kaya19}. For a review of these methods and their origin, see \cite{Chen17diag}. In Appendix \ref{app:CTP} we review the propagators involved in the Keldysh-Schwinger formulation of the CTP's diagrammatic expansion, paying attention to how they arise from the imposition of the initial state and the upper boundary of the closed-time-path at $t_{\rm f}$ which remains finite but arbitrary. 

As we commented above, $\gamma$ appears explicitly in the exponent of the evolution operator as it provides one of the first answers to the question 'what is quantum?'. In this expression $\gamma$ is a critical quantity for the interaction terms as the behaviour of the exponential strongly depends on it. However, dynamics are not the only key to a quantum theory. In particular, one of the key features of eq. \eqref{eq:ininqcl} is the cosmological and quantum imprint on the formula: it is assumed that there exists a time (finite or infinite) where the inflaton is in its quantum vacuum state (defined by its free Hamiltonian) and its evolution gives us the state of the inflating universe. The dynamics are highly dependent on how the initial conditions scale compared to $\gamma$. 

%It is often said that quantum corrections arise as powers of $\gamma$. 
A convenient way to distinguish quantum from classically reproducible dynamics lies in the so-called Keldysh formalism of the closed-path integral \cite{Keldysh64}, found with a simple field redefinition
\begin{equation}
\begin{pmatrix}
    {\cal R}_+ \\
    {\cal R}_-
\end{pmatrix}
=
\begin{pmatrix}
    1 & \frac{\gamma}{2} \\
    1 & -\frac{\gamma}{2}
\end{pmatrix}
\begin{pmatrix}
    {\cal R}_{cl} \\
    {\cal R}_{q}
\end{pmatrix}.
\label{eq:redefKeldysh}
\end{equation}
As we hinted to above,\textit{ $\gamma$ will be a parameter counting semi-classical orders in our development.} {In particular, as will be seen more explicitly further below, the basis change \eqref{eq:redefKeldysh} leads to a re-definition of propagators from ``$++$", ``$--$" and ``$+-$" types, directly arising from \eqref{eq:ininqcl}, to ``$\rm{cl}$$-$$\rm{q}$ and ``$\rm{cl}$$-$$\rm{cl}$" propagators - see \eqref{eq:defGKGR}. This will allow for a direct counting of powers of $\gamma$ and a corresponding ``classicality distinction" between diagrams with differing numbers of these propagators.} The basis change is usually performed on the functional path integral  \cite{Keldysh64,vanderMeulen2007, Radovskaya2021} but can also be performed directly in the \textit{in-in} formula. 

For example, when considering a simple toy $-\lambda a {\cal R}^3\subset {\cal H}_{int}$ interaction, the first order in-in bispectrum contribution at the ($t=t_f$) boundary is
\begin{equation}
\begin{aligned}
        \langle \hat{\cal R}_{\boldsymbol{k}_1} \hat{\cal R}_{\boldsymbol{k}_2}\hat{\cal R}_{\boldsymbol{k}_3}\rangle   %& = \langle 0 | \Bar{T}e^{-\frac{i}{\gamma}\lambda\int d^4x a(\hat{\cal R}_-^{\cal I})^3}  \hat{\cal R}_{\boldsymbol{k}_1} \hat{\cal R}_{\boldsymbol{k}_2}\hat{\cal R}_{\boldsymbol{k}_3} Te^{\frac{i}{\gamma}\lambda\int d^4xa(\hat{\cal R}_+^{\cal I})^3} | 0\rangle \\
        & \simeq -\frac{i}{\gamma}\lambda \int d^4x a \langle 0 |  \hat{\cal R}_-(x)^3  \hat{\cal R}_{\boldsymbol{k}_1} \hat{\cal R}_{\boldsymbol{k}_2}\hat{\cal R}_{\boldsymbol{k}_3}  -  \hat{\cal R}_{\boldsymbol{k}_1} \hat{\cal R}_{\boldsymbol{k}_2}\hat{\cal R}_{\boldsymbol{k}_3}\hat{\cal R}_+(x)^3  | 0\rangle+{\cal O}(\lambda^2),\\
\end{aligned}
\label{eq:exampleR3}
\end{equation}
which at this stage would usually be simplified as a commutator and the $\pm$ difference left in the $\pm i\epsilon$ prescription of time integrals. The decomposition of these interaction picture operators in the vacuum's anihilation and creation basis in Fourier space would then allow to keep only integrated contractions thanks to Wick's theorem. Using the $\pm$ propagators (defined in Appendix \ref{app:CTP}) for these and integrating over time will yield the full quantum correlation. This is the standard \textit{in-in} method.

The Keldysh method instead is adapted for use in the Dyson formula by substituting the decomposition of eq. (\ref{eq:redefKeldysh})  in eq. (\ref{eq:exampleR3}). In particular, at the ($t=t_f$) boundary where $\hat{\cal R} = \hat{\cal R}_{cl}$, the correlator can be expressed as
\begin{equation}
\resizebox{\textwidth}{!}{
$\begin{aligned}
  \langle \hat{\cal R}_{\boldsymbol{k}_1} \hat{\cal R}_{\boldsymbol{k}_2}\hat{\cal R}_{\boldsymbol{k}_3}\rangle = -\frac{i}{\gamma}\lambda \int_{k_a,k_b,k_c}\delta^{(3)}(\boldsymbol{k_a}+\boldsymbol{k_b}+\boldsymbol{k_c})\int_{t_{\rm i}}^{t_{\rm f}} dt \, a \langle 0 |  \prod_{\mu\in \{ a,b,c\}}\left (\hat{\cal R}_{cl,\boldsymbol{k}_{\mu}}-\frac{\gamma}{2}\hat{\cal R}_{q,\boldsymbol{k}_{\mu}}\right)  \hat{\cal R}_{cl,\boldsymbol{k}_1} \hat{\cal R}_{cl,\boldsymbol{k}_2}\hat{\cal R}_{cl,\boldsymbol{k}_3} | 0\rangle\\
  - a \langle 0 | \hat{\cal R}_{cl,\boldsymbol{k}_1} \hat{\cal R}_{cl,\boldsymbol{k}_2}\hat{\cal R}_{cl,\boldsymbol{k}_3}\prod_{\mu\in \{ a,b,c\}}\left (\hat{\cal R}_{cl,\boldsymbol{k}_{\mu}}+\frac{\gamma}{2}\hat{\cal R}_{q,\boldsymbol{k}_{\mu}}\right)  | 0\rangle+{\cal O}(\lambda^2),
\end{aligned}$
}
\label{eq:exampleR3bis}
\end{equation}
which greatly simplifies and over the tetrapyd becomes
\begin{equation}
\begin{aligned}
  \langle \hat{\cal R}_{\boldsymbol{k}_1} \hat{\cal R}_{\boldsymbol{k}_2}\hat{\cal R}_{\boldsymbol{k}_3}\rangle' = -2\lambda\int^{t_f}_{t_{\rm in}} dt \, a(t) K_{k_1}(t, t_f)K_{k_2}(t, t_f)G^R_{k_3}(t_f,t)\\
+\frac{1}{4}{\gamma^2}\lambda\int^{t_f}_{t_{\rm in}} dt a(t) G^R_{k_1}(t_f,t)G^R_{k_2}( t_f,t)G^R_{k_3}(t_f,t)+k-perm.+{\cal O}(\lambda^2)\,.\\
\end{aligned}
\label{eq:exampleR3ter}
\end{equation}
Here, we have used Wick's theorem, kept connected diagrams, and simplified terms with even powers of $\hat{\cal R}_{q}$,\footnote{The opposite signs come from having $i$ factors  in $G$ propagators but not $K$.} with the propagators in this basis defined as
\begin{equation}
\left \{
    \begin{aligned}
        iK(k,t,t') & \equiv \langle \hat{\cal R}_{cl,\boldsymbol{k}}(t)\hat{\cal R}_{cl,-\boldsymbol{k}}(t')\rangle \equiv  F(k,t,t'),   \\
        iG^R(k,t,t') & \equiv \langle \hat{\cal R}_{cl,\boldsymbol{k}}(t)\hat{\cal R}_{q,-\boldsymbol{k}}(t')\rangle  \equiv -{i}\theta[t-t']G(k,t,t'),  \\
        iG^A(k,t,t') & \equiv iG^R(k,t',t)\,.
    \end{aligned}\right .
    \label{eq:defGKGR}
\end{equation} 
These can be computed explicitly with the free theory and the initial conditions as shown in Appendix \ref{app:solveGreen}. Here we have parametrised the magnitudes via the Fourier transforms of the Schwinger (up to $-2i$ factor) and Hadamard functions respectively\footnote{Also known as real and imaginary parts of the Wightman function.}
\begin{equation}
\left \{
    \begin{aligned}
        F(k,t,t') &= Re[ {\cal R}_k(t) {\cal R}_k^*(t')],   \\
       G(k,t,t') & = 2Im[{\cal R}_k(t) {\cal R}_k^*(t')].  \\
    \end{aligned}\right .
    \label{eq:defFG}
\end{equation} 
Note that $K$ is not strictly speaking a Green function but a solution of ${\cal D}_0(t) K(t,t') = 0$ but is still labeled as a propagator. The crucial point to note about \eqref{eq:exampleR3ter} is that only the $\gamma^0$ terms can be reproduced by classical dynamics; the $\gamma^2$ terms are ``purely quantum'' and cannot be obtained via any sort of classical dynamics {augmented with statistical fluctuations}.

\begin{figure}[ht]
    \centering
    \begin{tabular}{cc}
        \scalebox{0.3}{
            \begin{feynman}
    \fermion[showArrow=false]{4.50, 7.00}{11.50, 7.00}
    \fermion[lineWidth=4pt]{8.00, 4.00}{11.00, 7.00}
     %\fermion{8.00, 4.00}{8.00, 7.00}
     \dashed[showArrow=false, lineWidth=5pt]{8.00, 4.00}{8.00, 7.00}
     \dashed[showArrow=false, lineWidth=5pt]{8.00, 4.00}{5.00, 7.00}
     \myvertexC[color=000000, lineWidth=2]{8.00, 4.00}{0.2}
          \draw (5.00, 7.4) node[scale=3.5] {$k_1$};
     \draw (8.00, 7.4) node[scale=3.5] {$k_2$};
     \draw (11.00, 7.4) node[scale=3.5] {$k_3$};
\end{feynman}
        } 
        &
        \scalebox{0.3}{
            \begin{feynman}
    \fermion[showArrow=false]{4.50, 7.00}{11.50, 7.00}
    \fermion[lineWidth=4pt]{8.00, 4.00}{11.00, 7.00}
     \fermion[lineWidth=4pt]{8.00, 4.00}{8.00, 7.00}
     \fermion[lineWidth=4pt]{8.00, 4.00}{5.00, 7.00}
     \myvertexQ[color=000000, lineWidth=2]{8.00, 4.00}{0.23}
     \draw (5.00, 7.4) node[scale=3.5] {$k_1$};
     \draw (8.00, 7.4) node[scale=3.5] {$k_2$};
     \draw (11.00, 7.4) node[scale=3.5] {$k_3$};
\end{feynman}
        } \\
        % Diagram 1 & Diagram 2 \\
    \end{tabular}
    \caption{Bulk-to-boundary tree-level diagrams from the $\lambda(t){\cal R}^3$ vertex {decomposed as Keldysh-Feynman `classical' (left) and `quantum' (right) diagrams, up to ($k_1$, $k_2$, $k_3$) permutations. A dashed line stand for a $K$ propagator while an oriented solid one stands for $G^{R/A}$ (see eq. \eqref{eq:defGKGR}). Vertex rules are explicit in eq. \eqref{eq:exampleR3ter}. The non-oriented solid line is the boundary (time $t_{\rm f}$), intersected at each field's value with momentum $k_i$.}}
    \label{fig:diag1}
\end{figure}
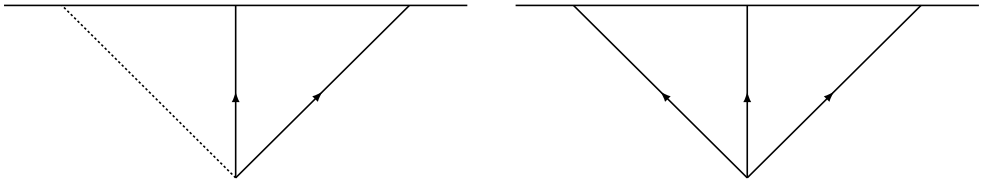

Using $\gamma$ as a counting parameter, one quickly understands that the interaction terms at any order will yield powers of $\gamma$ corresponding to factors of ${\cal R}_q$ in these expressions and in the interaction hamiltonian or action. Different powers of $\gamma$ yield different vertices. In general, only odd powers of ${\cal R}_q$ appear in the action of a closed system\footnote{An influence functional can arise when integrating out either an environment or a group of scales  \cite{Feynman63} and lead to a theory with interactions terms between $\pm$, and ultimately to non-odd powers of ${\cal R}_q$ in the resulting Keldysh action, see for instance the open EFT of Inflation \cite{Salcedo24}. This will be ignored until the special case of section \ref{sec:SI}. } and vertices in the Keldysh representation can only include $2n+1$ $\{n=0,1,2,\ldots\}$ factors of ${\cal R}_{q}$. In the ${\cal R}^3$ example studied here two vertices are allowed: one with one quantum ${\cal R}_{q}$ leg  and one with three. In our case, we are interested in the tree-level bulk-to-boundary correlators where ${\cal R}_q(t_{\rm f}) = 0$, which yields only two Feynman diagrams, displayed in Figure~\ref{fig:diag1}. When summing all these contributions with their permutations, one should find the same correlators as the usual in-in approach.

The diagrams with a single $G^{R/A}$ propagator (which have the lowest power in $\gamma$) are usually referred to as \textit{classical} ones, while others are designated as purely \textit{quantum}. A first way to see that is illustrated in Appendix \ref{app:CSA}. It can be summarised in the fact that ${\cal R}_{cl}$ undergoes on-shell dynamics (classical equations of motion) in the $\gamma \rightarrow 0$ limit. Note however that the initial conditions are unchanged. This limit is the semi-classical limit, the leading order of which is called the classical statistical approximation (CSA) \cite{Radovskaya2021} and equates indeed to using only one $G^{R/A}$ propagator per vertex.

Another way to understand the role played by the number of $K$ and $G^{R/A}$ propagators is to look at the commutators and {anti-commutators}
\begin{equation}
\left\{
\begin{aligned}
        \bigg\langle\left\{\hat{\cal R}(\mathbf{x}, t), \hat{\cal R}(\mathbf{y}, t^{\prime})\right\} \bigg\rangle & = 2\int d^3 k e^{i \mathbf{k} \cdot(\mathbf{x}-\mathbf{y})} \operatorname{Re}\left[{\cal R}_k(t) {\cal R}_k^*\left(t^{\prime}\right)\right] = \frac{F}{2},\\
        \bigg\langle\left[\hat{\cal R}(\mathbf{x}, t), \hat{\cal R}(\mathbf{y}, t^{\prime})\right]\bigg\rangle & =2\int d^3 k e^{i \mathbf{k} \cdot(\mathbf{x}-\mathbf{y})} \operatorname{Im}\left[{\cal R}_k(t) {\cal R}_k^*\left(t^{\prime}\right)\right] = G,
\end{aligned}\right .
\label{eq:WeinbergCom}
\end{equation}
In that sense, discarding multiple $G^{R/A}$ propagators reflects discarding the contributions of commutators to the dynamics. This link is particularly important to relate the Keldysh formalism in cosmology to previous studies of classicality which mostly focus on commutation \cite{Weinberg05, Weinberg06, Assassi13, Lim15, Martin16, burgess_minimal_2022}.

\subsection{Semi-classicality of inflation}
{From the above discussion, we arrive at the conclusion that full quantum results will be adequately approximated by using classical dynamics if vertices with higher powers of $\gamma$ provide negligible contributions to correlators. A given n-vertex $\lambda(t)\gamma^p{\cal R}_{cl}^{n-p}{\cal R}_q^p$, will contribute $p$ G functions and at most $n\,$$-$$\,p$ Keldysh ones ($F$) to the computation of any correlation function. This means that if 
\begin{equation}\label{eq:clasCond}
F\gg  G\,,
\end{equation}
than the dominant contributions will come from vertices with the lowest value of $p\,$$=$$\,1$. These contribute no powers of $\gamma$ (canceling with the $\gamma^{-1}$ in the exponent of the evolution operator, as can be seen in \eqref{eq:ininqcl}) and hence can be reproduced by fully classical dynamics. Therefore, \emph{when \eqref{eq:clasCond} holds, classical dynamics will provide an adequate description of the complete quantum dynamics}. This is a common condition known in non-equilibrium QFT frameworks \cite{berges2015}, which is reminiscent of the theorem implying the classicality of the full QFT from the classicality of the free linear theory, proved in \cite{LythSeery08}.} Counting powers of $G/F$ in calculations is therefore definitive and an important aim.

\begin{figure}[ht]
\centering
\includegraphics[width=0.65\textwidth]{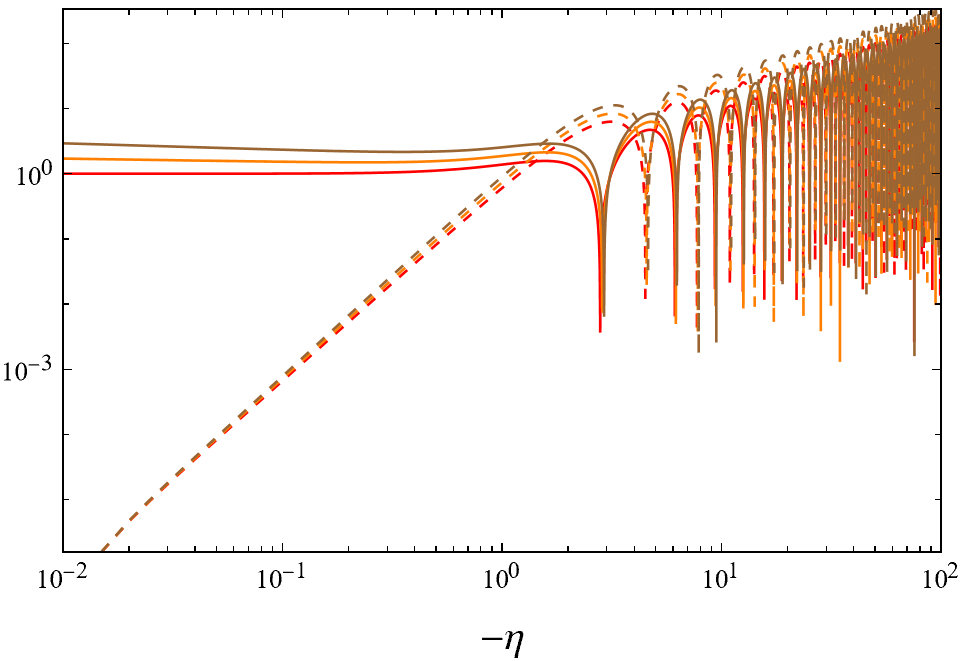}
    \caption{Absolute $F$ (solid lines) and $G$ (dashed lines) in quasi-dS ($\nu = \frac{3}{2}+0 \text{ (red) }, +0.05 \text{ (orange) }, +0.1 \text{ (brown) }$) for $\eta$ up to $\eta_f = -0.01 $ and $k=1$, up to constants. Oscillations amplify in the $\eta \rightarrow -\infty$ limit.}
    \label{fig:FG}
\end{figure}

Let us now focus on the present case provided by eqn. \eqref{eq:quasiDShankel} and its approximation eq. \eqref{eq:quasiDS}. One might have noticed that we purposefully did not set $\gamma = 1$ units as opposed to standard treatments. This is to emphasize that both on-shell and off-shell fields can have a $\gamma$ magnitude from their initial conditions. The corresponding kernels of eq. \eqref{eq:defFG} in Fourier space and conformal time become  \cite{vanderMeulen2007}
\begin{equation}
\left \{
    \begin{aligned}
& F\left(k, \eta_1, \eta_2\right)=\frac{\gamma H^2}{4 \varepsilon_1 M_{Pl}^2k^3}\left[\left(1+k^2 \eta_1 \eta_2\right) \cos k\left(\eta_1-\eta_2\right)+k\left(\eta_1-\eta_2\right) \sin k\left(\eta_1-\eta_2\right)\right], \\
& G\left(k, \eta_1, \eta_2\right)=\frac{\gamma H^2}{ 2\varepsilon_1 M_{Pl}^2k^3}\left[\left(1+k^2 \eta_1 \eta_2\right) \sin k\left(\eta_1-\eta_2\right)-k\left(\eta_1-\eta_2\right) \cos k\left(\eta_1-\eta_2\right)\right]\,.
\end{aligned}\right .
\label{eq:FGdS}
\end{equation}
These two kernels are plotted in Figure \ref{fig:FG}, together with quasi-dS solutions provided by the massive Hankel  solutions of eq. \eqref{eq:quasiDShankel}. For a given scale, it takes a certain time for the $F/G$ ratio to become effectively small, partially because UV scales have a divergent 2-point function but mostly because $F$ and $G$ are simply dephased and of comparable magnitude. For this class of solutions, one can show that $G$ goes to $0$ just after horizon crossing, while $F$ does not as shown in Figure \ref{fig:FG}. In that sense, discarding powers of $\gamma$ is allowed in quasi-dS if we focus on super-Hubble scales.

As we have just seen that inflationary semi-classicality can only be applied to a (time-dependent) subset of scales, one can wonder about how to adapt the formalism to the coarse-grained system. Are the propagators much impacted by the coarse-graining? Let us keep that aside for now as this caveat will be addressed in section \ref{sec:SI}.

\subsection{Inflationary calculations \label{sec:calc}}
In this section we ignore the non-validity of the semi-classical approximation (and so of the CSA) and study the on-shell and off-shell contributions to the bispectrum. We focus on the following interaction vertices:
\begin{equation}
\left \{
\begin{array}{cl}
H_{int} & = \displaystyle\int d^3x \sum_i {\cal H}^{(3)}_i \\
{\cal H}^{(3)}_i & = g_i(t)a(t)^{q_i}f_i[{\cal R}]
\label{eq:eqH3}
\end{array}\right .
\end{equation}
% \begin{equation}
% \begin{array}{cl}
% H_{int} = &\displaystyle\int ??\dot{\cal R}^3 -a^3\epsilon ^2 {\cal R} \dot{{\cal R}}^2 -a\epsilon ^2 {\cal R} (\partial{\cal R})^2 +2a\epsilon \dot{{\cal R}} (\partial{\cal R})(\partial\chi)\\
% & -\frac{\epsilon}{2a}(\partial_i{\cal R})(\partial^i\chi)\partial^2\chi-\frac{\epsilon}{4a}\partial^2{\cal R}(\partial\chi)^2  d^3x,
% \label{eq:eqH3}
% \end{array}
% \end{equation}
%
% where $\partial^2\chi = \frac{a^2\varepsilon_1\dot{\cal R}}{c_s^2}$, 
where the chosen couplings ($g_i$), scaling ($q_i$), and cubic differential functionals ($f_i$) are defined in Table \ref{tab:interactions}. This corresponds to leading cubic terms of an inflationary theory in standard General Relativity \cite{chen_observational_2007}. This means that we need to study vertices like $g_1(t)\dot{{\cal R}}^3$, $g_2(t){\cal R} \dot{{\cal R}}^2$, $g_3(t){\cal R}(\partial{\cal R})^2$, and $g_4(t)\dot{{\cal R}}\partial{\cal R}\partial\chi$, absorbing EOM terms by redefinition \cite{Maldacena_2003} (see Appendix \ref{app:appRedef}).

\renewcommand{\arraystretch}{1.5}
\begin{table}[ht]
\centering
\resizebox{\textwidth}{!}{%
\begin{tabular}{|c|c|c||c|c|}
\hline
$g_i(t)$ & $q_i$ & $f_i[{\cal  R}]$ & $f_i^q[{\cal  R}]$ & $f_i^{cl}[{\cal  R}]$\\
\hline
 $ H^{-3}(\Sigma(1-\frac{1}{c_s^2})+2\lambda)$ & $3$& $\dot{\cal R}_{\pm}^3$ & $\frac{1}{4}\dot{\cal R}_q^3$ & $2\dot{\cal R}_q\dot{\cal R}_{cl}^2$ \\
\hline
 $-\frac{\varepsilon_1}{c_s^4}(\varepsilon_1-3+3c_s^2)$ & $3$& ${\cal R}_{\pm}\dot{\cal R}_{\pm}^2$ & $\frac{1}{4}{\cal R}_q\dot{\cal R}_q^2$ &  $ {\cal R}_q\dot{\cal R}_{cl}^2+2 {\cal R}_{cl}\dot{\cal R}_{cl} \dot{\cal R}_q$  \\
\hline
 $-\frac{\varepsilon_1}{c_s^2}(\varepsilon_1-2s+1-c_s^2)$ & $1$& ${\cal R}_{\pm}(\partial{\cal R}_{\pm})^2$& $\frac{1}{4}{\cal R}_q(\partial{\cal R}_q)^2$ &  $ {\cal R}_q(\partial {\cal R}_{cl})^2+2 {\cal R}_{cl}\partial {\cal R}_{cl} \partial{\cal R}_q$\\
\hline
 $\varepsilon_1^2/c_s^4$ & $3$& $\dot{\cal R}_{\pm}(\partial{\cal R}_{\pm})(\partial\partial^{-2}\dot{\cal R}_{\pm})$ & $\frac{1}{4}\dot{\cal R}_q\partial{\cal R}_q\partial\partial^{-2}\dot{\cal R}_q$ & $\dot{\cal R}_q\partial{\cal R}_q\partial\partial^{-2}\dot{\cal R}_{cl}$ + 5 perm. \\
\hline
\end{tabular}}
\caption{Terms of the slow-roll leading cubic Hamiltonian density of GR coupled to a $P(X,\phi)$ inflaton after redefinition (see eq. (4.31) and notations of \cite{chen_observational_2007}) and associated Keldysh vertices after basis change.\label{tab:interactions}}
\end{table}
\renewcommand{\arraystretch}{1}
In \cite{vanderMeulen2007}, a $g \phi^3$ theory is studied in dS to derive the 2-point correlator at second order in $g$, including regulated loops. For a different purpose in another study \cite{Salcedo24}, vertex coefficients of an open EFT in the Keldysh basis have been looked at and bispectrum integral expressions have been found for (quasi-)dS spacetimes, and solved for numerically. In this work, we managed to perform analytical integrations by using the quasi-dS solutions of eq. \eqref{eq:quasiDS} in the integrands and simplifying the results using \textit{Mathematica}. In the following, we use the elementary symmetric polynomials to compactify the writing of momenta mixings
\begin{equation}
    \left\{\begin{aligned}
        e_1 & = k_1+k_2+k_3, \\
        e_2 & = k_1k_2+k_2k_3+k_1k_3, \\
        e_3 & = k_1k_2k_3.
    \end{aligned} \right .
\end{equation}
For the sake of generality, our calculations have not been restricted to correlators at the conformal boundary ($\eta = 0$) but calculated for any future boundary $\eta\leq 0$. Similarly to usual computations of the literature \cite{chen_observational_2007}, we make the following assumptions for these derivations:
\begin{itemize}
    \item[\ding{118}] the integrations are carried from the far past ($t_{\rm in}$ or $\eta_i = -\infty$) to the boundary, with the right $\pm i\epsilon$ prescription in any integration of diverging oscillations in the $t_{\rm in}$ limit: the prescription is still necessary in this basis.
    \item[\ding{118}] propagators (such as $K$, $G^{A/R}$, or $G^{\pm\pm}$) can be approximated by their quasi-dS form of eq. (\ref{eq:FGdS}).
    \item[\ding{118}] coupling parameters ($\lambda, \Sigma, \varepsilon_1,\varepsilon_2,H$) are assumed to be slowly varying so that they can be taken out of the time integrals. For these reasons, we will set these almost constants to $1$ and focus on the $k_i$ dependence. For instance eq. \eqref{eq:FGdS} will be used with $\frac{\gamma H^2}{ 4\varepsilon_1 M_{Pl}^2k^3} \rightarrow 1$.
    \item[\ding{118}] we are not interested in the redefinition contribution one would typically encounter to simplify EOM terms \cite{Maldacena_2003, chen_observational_2007}. These generate products of 2-point functions and are thus essential to find the consistency relations for instance \cite{Maldacena_2003}. Separating the quantum and classical part of these at first order in the coupling is straightforward because it is just the free theory. On top of that, evaluating the correlators at the boundary $\eta_f = \eta$ removes the ${\cal R}_q$ field, which means it is a purely classical effect at this order.
\end{itemize}
In practice, our code decomposes the integrals into sums of Euler $\Gamma$ functions (regularised by the $\pm i \epsilon$ prescription), which have analytical expressions. We give the final results below, including Fourier space permutations and keeping the tetrapyd's delta functions implicit. In the following, ${\cal B}_{i} =  \langle \hat{\cal R}_{\boldsymbol{k}_1} \hat{\cal R}_{\boldsymbol{k}_2}\hat{\cal R}_{\boldsymbol{k}_3}\rangle'$ for the type $i$ interaction vertex using \textit{in-in} formalism. Inspired by eq. (\ref{eq:exampleR3ter}), ${\cal B}^{cl}_{i}$ (resp. ${\cal B}^{q}_{i}$) are the $FFG$ (resp. $GGG$) contributions only, calculated with the Keldysh method and giving the on-shell (resp. off-shell) interactions. 
For $a^3\dot{\cal R}^3$ and at final conformal time $\eta\leq 0$, we get
\begin{equation}
\left \{
\begin{aligned}
{\cal B}_{1} & ~\propto~  \frac{6 e_1 e_3 \eta ^4+6 e_2 \eta ^2-3 e_1^2 \left(e_2 \eta ^4+\eta ^2\right)-6}{2 e_1^3 e_3}, \\
 {\cal B}^{q}_{1} & ~\propto~  3 \left(3 e_1^6-18 e_1^4 e_2+24 e_1^3 e_3+24 e_1^2 e_2^2-48 e_1 e_2 e_3+32 e_3^2\right)\\
 & ~~~~~\times \frac{ \left(e_1^4 \eta ^4-4 e_1^2 \left(e_2 \eta ^4+\eta ^2\right)+8 e_1 e_3 \eta ^4+8 e_2 \eta ^2-8\right)}{2 e_1^3 \left(e_1^3-4 e_1 e_2+8 e_3\right)^3}.\\
\end{aligned}\right .
\label{eq:dot3}
\end{equation}
For $a^3{\cal R} \dot{{\cal R}}^2$ we get
\begin{equation}
\left \{
\begin{aligned}
   {\cal B}_{2} & ~\propto~  \frac{e_1^2 \left(3\eta^2 e_2-2\right) e_3+e_2 \left(1-\eta ^2 e_2\right) e_3+e_1e_2 \left(-e_3^2 \eta ^4-e_2^2 \eta ^2+e_2\right)}{2 e_1^2 e_3^3}, \\
    {\cal B}^{q}_{2} & ~\propto~  \frac{\left(e_1^2-2 e_2\right) \left(\eta ^2 \left(8 e_2+e_1 \left(e_1 \left(\eta ^2 \left(e_1^2-4 e_2\right)-4\right)+8\eta^2 e_3\right)\right)-8\right)}{2 e_1^2 \left(e_1^3-4e_1 e_2+8 e_3\right){}^2}.\\
\end{aligned} \right .
\label{eq:dot2}
\end{equation}
For $a{\cal R} (\partial {\cal R})^2$ we get
\begin{equation}
\left \{
\begin{aligned}
   {\cal B}_{3} & ~\propto~ \frac{\left(e_1^2-2 e_2\right) \left(e_1^3+\left(e_3^2 \eta ^4+e_2^2 \eta ^2-e_2\right) e_1-2e_1^2 e_3 \eta ^2+e_3 \left(e_2 \eta ^2-1\right)\right)}{4e_1^2 e_3^3}, \\
{\cal B}^{q}_{3} & ~\propto~ -\frac{\left(e_1^2-2 e_2\right) \left(8+4 \left(e_1^2-2 e_2\right) \eta ^2 -e_1 \left(e_1^3-4e_1 e_2+8 e_3\right) \eta ^4 \right)}{2e_1^2 \left(e_1^3-4e_1 e_2+8 e_3\right)^2}.    \\
    \end{aligned}\right .
    \label{eq:grad2}
\end{equation}
For $a^3\dot{\cal R} (\partial {\cal R})(\partial \partial^{-2}\dot{\cal R})$ we get
\begin{equation}
\left \{
\begin{aligned}
    {\cal B}_{4}& ~\propto~ \frac{-2 e_1^5+7 e_1^3 e_2+e_1 e_3 \eta ^4 \left(e_3 \left(e_1^2+4 e_2\right)+e_1 e_2 \left(e_1^2-4 e_2\right)\right)-17 e_1^2 e_3}{8 e_1^2 e_3^3}\\
    &~~~~~~+\frac{\eta ^2 \left(e_1^5 e_2-e_1^4 e_3-3 e_1^3 e_2^2+13 e_1^2 e_2 e_3-4 e_1 e_2^3+4 e_2^2 e_3\right)+4 e_1 e_2^2-4 e_2 e_3}{8 e_1^2 e_3^3},\\
{\cal B}^{q}_{4} & ~\propto~ \left(-e_1^2 e_2^2 \left(e_1^2-4 e_2\right)+2 e_3^2 \left(7 e_1^2+2 e_2\right)+2 e_1 e_3 \left(e_1^4-4 e_1^2 e_2-4 e_2^2\right)\right) \\
& ~~~~~\times\frac{ \left(e_1 \eta ^4 \left(e_1^3-4 e_1 e_2+8 e_3\right)-4 \eta ^2 \left(e_1^2-2 e_2\right)-8\right)}{4 e_1^2 e_3^2 \left(e_1^3-4 e_1 e_2+8 e_3\right)^2}. \\
    % {\cal B}^{cl}_{4} \propto& \\
    \end{aligned}\right .
    \label{eq:invdot2}
\end{equation}
Note that the classical bispectra ${\cal B}_i^{cl}$ have also been computed independently with their permutations and all satisfy ${\cal B}_i= {\cal B}_i^{cl}+{\cal B}_i^{q}$.
These are also all in agreement with eqns. (4.33-4.36) of \cite{chen_observational_2007} in the $\eta = 0$ limit with its unit-full coefficients. Another check is that of \cite{Salcedo24}, conjecturing the folded singularities we encounter here and similarly in \cite{Green20} where the first interaction's classical vertex contribution was computed in the $\eta = 0$ limit and matches eq. (\ref{eq:dot3}).

Interesting observations can be made when comparing the off-shell and on-shell contributions.
We start by taking the \textbf{squeezed limit} in those expressions, which makes $e_1$ and $e_2$ functions of two momenta only, while $e_3 \longrightarrow 0$. In all these expressions, we find 
    \begin{equation}
            {\cal B}_{i} \sim {\cal B}^{cl}_{i} \gg
           {\cal B}^{q}_{i}.
        \label{eq:squeezed}
    \end{equation}
{In this limit, the quantum contribution is ${\cal O}(1)$ (constant of $e_1$ and $e_2$), while the classical parts diverge as $e_3^{-3}$ or $e_3^{-1}$ (interaction $1$).
The result is interesting when thinking of the consistency relation verified by the squeezed limit of the single-field 3-point function \cite{Maldacena_2003, Creminelli04, Ganc10,RenauxPetel10}}
{
\begin{equation} \lim_{\substack{k_3 \rightarrow0}} {\cal B}(k_1,k_2,k_3) = (1-n_s)P_{k_3}P_{k_1}, \label{eq:consistency}\end{equation}
where $P_k = (2\pi)^{-3}\langle \hat{\cal R }_{\vec{k}}\hat{\cal R }_{-\vec{k}}\rangle$ and $n_s-1 = \frac{d \ln [k^3P_k]}{d \ln k}|_{k_1}$, and since $k_2\simeq k_1$ in this limit. This is true for the previous full \textit{in-in} computations in eqns. \eqref{eq:dot3}, \eqref{eq:dot2}, \eqref{eq:grad2}, and \eqref{eq:invdot2}.  In particular, all of the mentioned derivations rely on the classicality of at least one long wavelength mode (super-Hubble $k_3$) and suggest that the squeezed limit is the result of the free quantum noise evolution on the perturbed classical background.  This explains why, among other things, a more rigorous $\Delta N$ formalism computation can recover the same result \cite{Abolhasani_2018}. The domination of the squeezed limit by the on-shell evolution expressed by our result in eq. \eqref{eq:squeezed} adds an element of depth because interactive classicality has been well defined and identified. Of course, one should not forget that redefinition terms in Appendix \ref{app:appRedef} contribute to the consistency relation too. These are much easier to handle in the long wavelength regime where ${\cal R}$ freezes and small gradients cancel terms. In this limit or not, it is appropriate to assume that these are on-shell at leading order in the couplings, which implies that eq. \eqref{eq:consistency} is completely explained by on-shell evolution, at least at leading order.}

In the \textbf{equilateral limit}, there is nothing special appearing at first sight which would make either classical or quantum contributions dominate. However, the \textbf{folded limit} (i.e. $k_1 = k_2+k_3$ and cyclic permutations) is more relevant. It is dominated by the folded  singularities of the form 
    $$\left(e_1^3-4e_1 e_2+8 e_3\right)^2 = (k_1+k_2-k_3)^2 (k_1-k_2+k_3)^2(-k_1+k_2+k_3)^2$$
    which only appear when using the classical and quantum Keldysh contributions, as opposed to the total \textit{in-in} results. In such a configuration, the leading order $\left(e_1^3-4e_1 e_2+8 e_3\right)^{-2}$ of those singularities is
    \begin{equation}
        {\cal B}^{cl}_{i}  \sim - {\cal B}^{q}_{i} \gg {\cal B}_{i}.
    \end{equation}
    We see that the folded limit sees leading contributions cancelling between the on-shell and off-shell parts, so that the total bispectrum $ {\cal B}_{i}$ is finite. Note that accounting for dissipation effects can change the conclusion by smoothing out the folded singularities \cite{Salcedo24}. In \cite{Green20}, one can find an interpretation for these emerging singularities. While the fully quantum \textit{in-in} contributions do not exhibit any physical poles because of the uncertainty principle (i.e. no positive energy for virtual particles of the vacuum), the (on-shell) classical evolution does produce physical particles and so physical poles.

    Overall, it is clear that the shape plays a crucial role in deciding whether a 3-point function can be computed via a fully classical evolution, at least from an infinite past. Only squeezed enough shapes seem to be fully reproducible with such classical evolution of initial quantum fluctuations, though later we uncover some notable exceptions to these expectations numerically.  
    
\section{Timing classicality \label{sec:sec2}}
\subsection{\textit{In-in} truncations \label{sec:finite}}
\begin{figure}[ht]
    \centering
    
\includegraphics[width=0.87\textwidth]{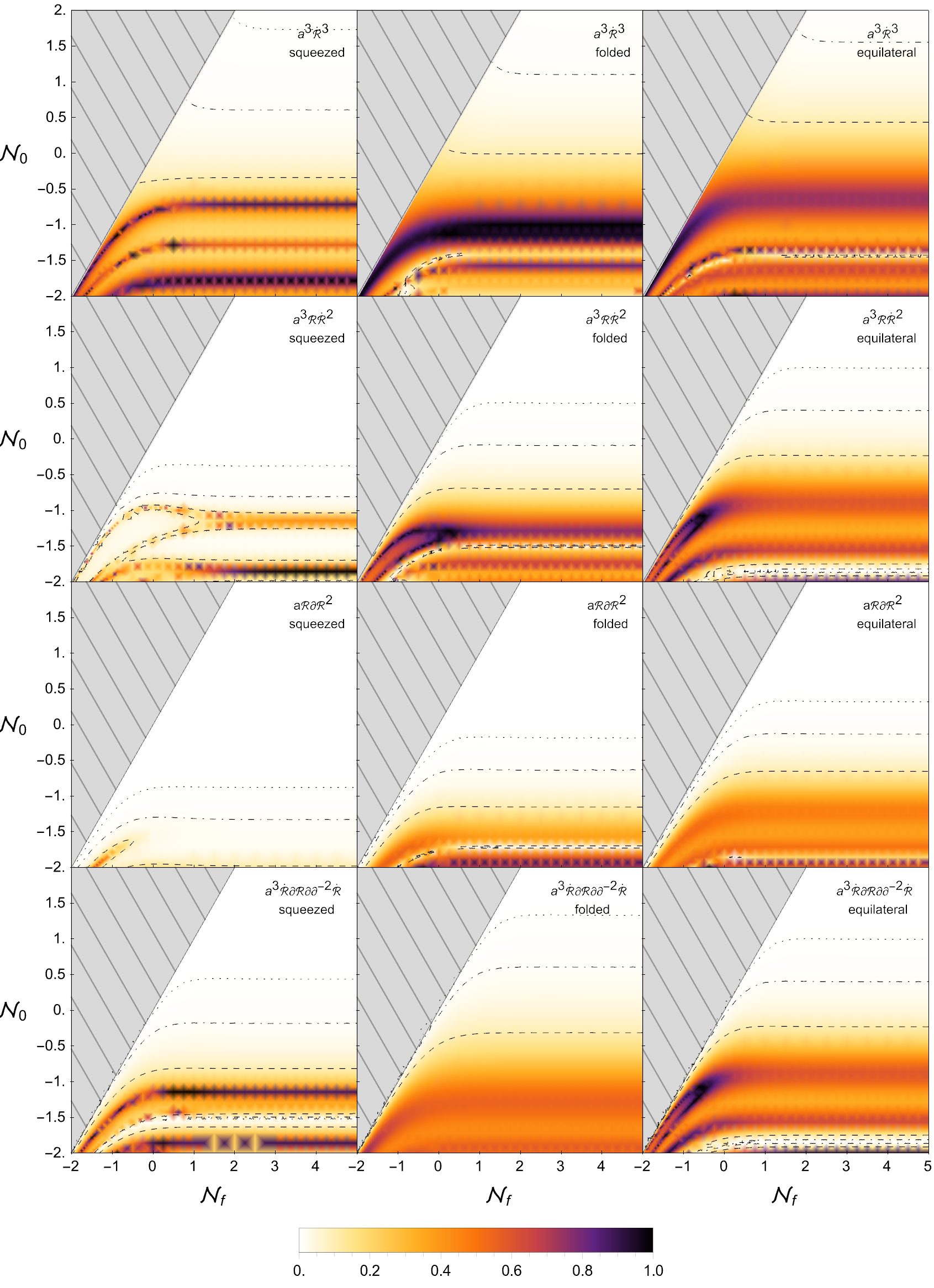}
    \caption{$ {\cal QI}(t_0,t_f)$ for $g_1 a^3\dot{{\cal R}}^3$, $g_2 a^3{\cal R} \dot{{\cal R}}^2$, $g_3 a{\cal R}(\partial{\cal R})^2$, and $g_4a^3\dot{\cal R}(\partial{\cal R})(\partial\partial^{-2}\dot{\cal R})$ in quasi-dS (top to bottom) for {squeezed ($k_1 = k_2 = k_3/0.1 = 1$), folded ($k_1 = k_2/0.55 = k_3/0.5 = 1$), and equilateral ($k_1 = k_2 = k_3 = 1$)} limits (left to right). $k_1$ is set to $1$ and is the last mode to cross the horizon ($k_{2,3}\leq k_1$), time at which we set the time origin ($k_1=aH$ at $\eta = -1$ or ${\cal N} = 0$  equivalently). Hatched grey regions stand for non-causal regions. Dashed contours have values 0.1 (dashed) and 0.01 (dashed dotted) and 0.001 (dotted). %Conversion to conformal times reads $\eta = -e^{-\cal N}/k_1 = -7.389, -2.718, -1.000, -0.368 ,-0.135, -0.050, -0.018,-0.007 $ for ${\cal N} = -2, -1, ..., 5$. 
    }
    \label{fig:RatioTime}
\end{figure}

\begin{figure}[ht]
    \centering
     \hspace*{-0.12\textwidth} 
\includegraphics[width=1.2\textwidth]{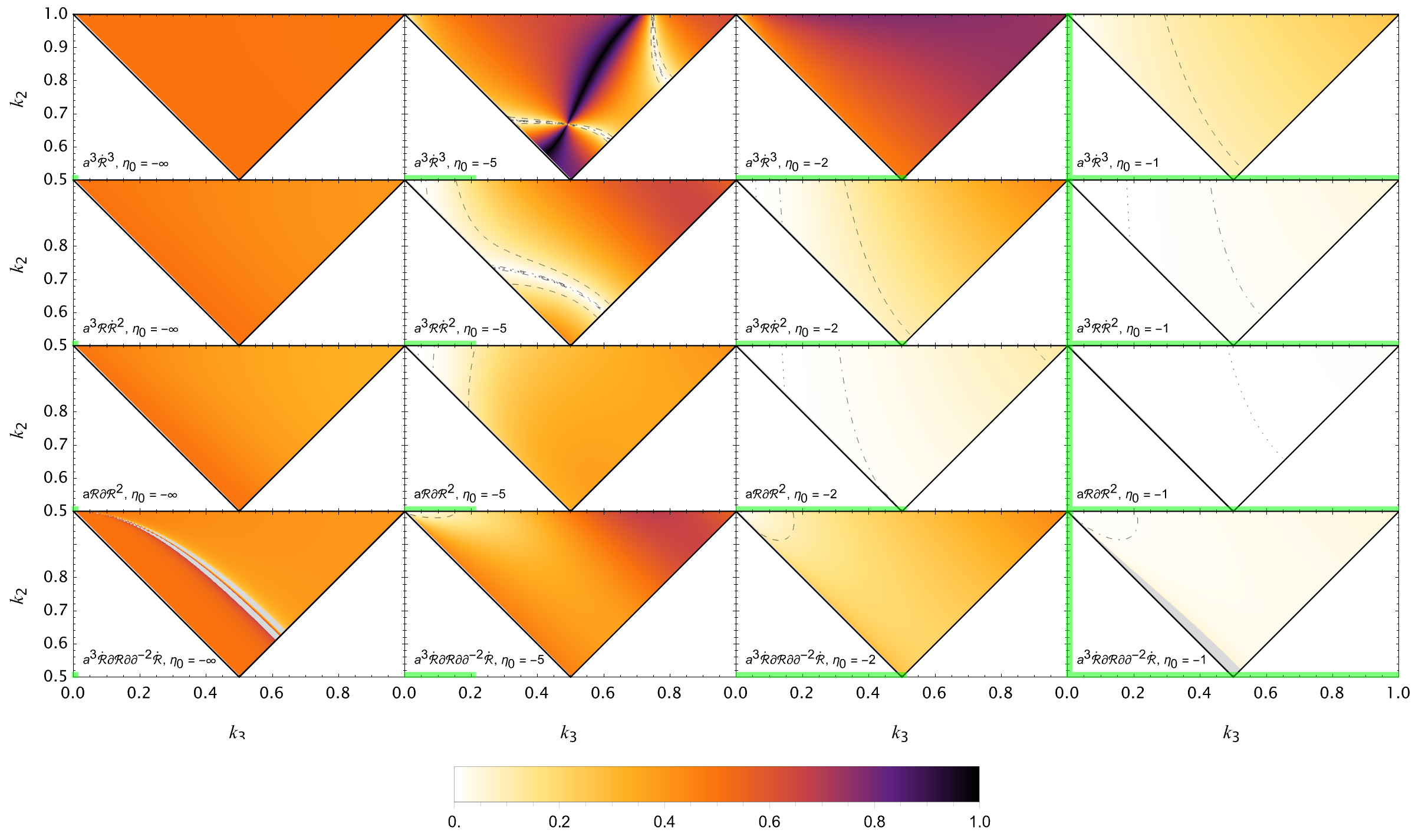}
    \caption{$ {\cal QI}(t_0,t_f) (1,k_2,k_3,\eta_0, \eta_f = -0.01)$ for $g_1 a^3\dot{{\cal R}}^3$, $g_2 a^3{\cal R} \dot{{\cal R}}^2$, $g_3 a{\cal R}(\partial{\cal R})^2$, and $g_4a^3\dot{\cal R}(\partial{\cal R})(\partial\partial^{-2}\dot{\cal R})$ in quasi-dS (top to bottom) and $\eta_0 =  -\infty^{\pm}, -5, -2, -1$ (left to right, equivalent to ${\cal N}_0 \simeq  -\infty, -1.61, -0.69, 0
    $). Range from $0$ (white) to $1$ (black) before reaching plot threshold or numerically indeterminate (grey). Dashed contours have values 0.1 (dashed) and 0.01 (dashed dotted) and 0.001 (dotted). Green gauges indicated which part of the mode ticks have crossed the horizon. $k_1$ is set to $1$ and is the last mode to cross the horizon ($k_{2,3}\leq k_1$), time at which we set the time origin ($k_1=aH$ at $\eta = -1$ or ${\cal N} = 0$  equivalently).}
    \label{fig:RatioShape}
\end{figure}

We have just argued that we can reliably compute the stochastic evolution of the bispectrum from the Bunch-Davies vacuum in the infinite past and that it is not satisfactory to neglect quantum interactions if QFTCS, together with the BD initialisation, is the appropriate theory describing our universe. We have also seen that classicality can be probed by considering the ratio of the two-point Keldysh and Green propagators. It is true that classicality beyond free theory is implied by that of the free theory itself \cite{LythSeery08} but we are interested in knowing precisely when it can be a good approximation to neglect quantum effects up to a desired accuracy in the $n$-point correlators. As one might expect, the answer is both shape and vertex-dependent. To study this issue, we cut the previous integrals into early and late-time quantum or classical contributions
\begin{equation}
\begin{aligned}
\langle {\cal O}(\boldsymbol{k}_1,...,\boldsymbol{k}_n,\eta)\rangle' & = {\cal C}_{early}+{\cal C}_{late} \\
& = {\cal C}_{early}^{q}+{\cal C}_{late}^{q} + {\cal C}_{early}^{cl}+{\cal C}_{late}^{cl},
\end{aligned}
\end{equation}
where the right early-late separation we seek satisfies ${\cal C}_{late}^{q}\ll{\cal C}_{late}^{cl}$ for given $\{\boldsymbol{k}_i\}$ at a time $\eta$.
At first order, the obvious separation is that of the time integration. One can truncate the previous integrals (such as eq. \eqref{eq:exampleR3ter}) to define schematically for any couplings:
\begin{equation}
\left\{
    \begin{aligned}
        {\cal C}_{early}^{cl} & = \sum_{(a,b,c)\in \text{ perm}(3) }\kappa_{abc}\left (\int_{-\infty}^{t_0} dt\theta_aF_{k_1}\theta_bF_{k_2}\theta_cG_{k_3}|_{(t,t_f)} +k\text{-perm.}\right )\,,\\
        {\cal C}_{late}^{cl} &  = \sum_{(a,b,c)\in \text{ perm}(3) }\kappa_{abc}\left (\int_{t_0}^{t_f} dt\theta_aF_{k_1}\theta_bF_{k_2}\theta_cG_{k_3}|_{(t,t_f)} +k\text{-perm.}\right )\,,\\
        {\cal C}_{early}^{q} &  = \sum_{(a,b,c)\in \text{ perm}(3) }-\kappa_{abc}\left (\int_{-\infty}^{t_0} dt\theta_aG_{k_1}\theta_bG_{k_2}\theta_cG_{k_3}|_{(t,t_f)} +k\text{-perm.}\right )\,,\\
        {\cal C}_{late}^{q} &  = \sum_{(a,b,c)\in \text{ perm}(3) }-\kappa_{abc}\left (\int_{t_0}^{t_f} dt\theta_aG_{k_1}\theta_bG_{k_2}\theta_cG_{k_3}|_{(t,t_f)} +k\text{-perm.}\right )\,,\\ 
    \end{aligned}\right .
\end{equation}
{where the $\theta_i$ are linear differential operators from the cubic interaction $\theta_1 {\cal R}\theta_2{\cal R}\theta_3 {\cal R}\subset S^{(3)}$, and $\kappa$ designates the Keldysh coefficients arising from the basis change (vertex-dependent,  see Table \ref{tab:interactions})}.
Note that early and late contributions are not the same because of the different integration range. If one wants to calculate explicitly these expressions with the same approximations as in the last section, it is recommended to proceed numerically (e.g.\ with \textit{Mathematica}) because of the many (oscillating) terms arising from such a truncation or shift in the range (which already appear if doing so with the total \textit{in-in} integrals). The cumbersome analytical expressions can be found thanks to our code, which can be provided on request. In this article, we simply show the outcomes graphically in Figures \ref{fig:RatioTime} and \ref{fig:RatioShape}, showcasing the key ratio which we refer to as the \textit{quantum interactivity} (QI):
\begin{equation}
    {\cal QI}(t_0,t_f,\{k_j\}) = \frac{|{\cal C}^q_{late}(t_0,t_f,\{k_j\}) |}{|{\cal C}^{cl}_{late}(t_0,t_f,\{k_j\}) |+|{\cal C}^q_{late}(t_0,t_f,\{k_j\}) |}\,,
\end{equation}
which can be generalised to any $n$-point function, such as the trispectrum, but also to multiple interactions and to each order in the coupling. It peaks at $1$ when the quantum contributions dominate fully or falls to $0$ in the opposite case. Note that the denominator is not the summed \textit{in-in} late contribution because we take the absolute value, preventing cancellation.

In Figure \ref{fig:RatioTime}, we look at the quantum interactivity for each cubic term (or shape) while varying $t_0$ and $t_f\geq t_0$ in the e-folding time ${\cal N}$. The e-folding time always has its origin set at the crossing of the largest $k$ (here $k_1=1$) because this represents the last mode to cross outside the horizon.  Overall trends In Figure \ref{fig:RatioTime} are noted first. For each fixed boundary ${\cal N}_f$, looking at a range in ${\cal N}_0$ allows us to see when the interactive quantum contributions are below $10\%$, $1\%$ or $0.1\%$ of the total contribution. 
On the other hand, the range in ${\cal N}_f$ tells about the stability in time of this behaviour. In particular, all these plots have in common the stability of ${\cal QI}$ after a few e-folds for each given ${\cal N}_0$, which is due to the suppression of the $G$ propagators in the integrands for both types of contributions.

Furthermore, these plots also show a hierarchy in `classicality' between interactions and their associated shapes. Again in Figure \ref{fig:RatioTime}, observe that the lower the dashed contours, the earlier one can start to neglect quantum contributions. From this perspective, it is clear that for the three shapes under consideration, the last interaction to become classical is $g_1(t)\dot{{\cal R}}^3$, preceded by $g_2(t){\cal R} \dot{{\cal R}}^2$ and $g_4(t)\dot{{\cal R}}\partial{\cal R}\partial\chi$, with the earliest being $g_3(t){\cal R}(\partial{\cal R})^2$. It thus appears that the more time derivatives, the later the classicalisation of the resulting bispectrum contribution. For each interaction, there is a close correspondence to the shape, from which we infer that there is more time needed for the equilateral shape compared to the folded one and the (mildly) squeezed one. A properly squeezed limit would confirm our conclusion from the previous section, which is that the local shape is mostly evolving classically.

Figure \ref{fig:RatioShape} illustrates the quantum interactivity on a slice through the full tetrapyd over a fixed ${\cal N}_0$--${\cal N}_f$ integration range. In particular, this provides confirmation of the hierarchy in classicality between the different interactions. Most importantly,  it confirms that the equilateral limit is typically the slowest to classicalise and thus constitutes a good diagnostic for when the full tetrapyd will have attained classicality. 
This figure illustrates again that the non-canonical $g_1(t)\dot{\cal R}^3$ interaction receives quantum contributions across a wider range of configurations until they are much closer to horizon crossing than the other terms.

A remarkable observation is that the ${\cal QI}$ diagnostic indicates that sub-Hubble evaluations can provide a good approximation to the final bispectrum for most interactions (where this can be slightly to considerable depending on shape). The possibility of sub-Hubble `classicality' is certainly not apparent from looking at the ratio $(F/G)^2$ (see Figure \ref{fig:FG}, which clearly takes $1$ to $2$ efolds after horizon crossing to reach a comparable level. This means that both the integration, the vertex and Keldysh coefficients (including $a(t)$) are important for calculating and identifying when the quantum to classical transition occurs. Performing the interactive calculus with our approximations is thus useful and quantitative, but one can also obtain excellent diagnostics for classicality at arbitrary accuracy using fully numerical solutions for specific models.\footnote{This is left for future work \cite{LRSZ}, using recent advances in fast bispectral predictions \cite{clarke_probing_2021}.} The question that is being addressed becomes: how long does it take for the cumulative $K$ propagator contributions to take over those from the $G$ propagators, before the latter closes down any further cumulative additions to the overall integral.

We conclude that, from the point of view of interactions, classicality can be reached earlier than previous criteria had suggested for most interactions. However, we note that the answer will always be interaction and shape dependent. In a way, these results are unprecedented but they are not actually different from what a quantitative study of commutator contributions would lead to when following other studies in the literature, had these been calculated more explicitly. For example, similar to the link established by eq.~(\ref{eq:WeinbergCom}) in free theory, eq.~(14) given in ref.~\cite{Green20} shows this in interactive theory:
\begin{equation}
\begin{aligned}
     {\cal B}_i(\boldsymbol{x_1},\boldsymbol{x_2},\boldsymbol{x_3},t)-{\cal B}_i^{cl}(\boldsymbol{x_1},\boldsymbol{x_2},\boldsymbol{x_3},t) ~\propto~~ ig_i \int^{t}_{t_0} dtd^3\boldsymbol{x} a^{q_i}\langle\theta^i_{1\boldsymbol{x}}\left [ \hat{\cal R}(\boldsymbol{x_1},t),\hat{\cal R}(\boldsymbol{x},t)\right] \\
     \times\,\theta^i_{2\boldsymbol{x}}\left [ \hat{\cal R}(\boldsymbol{x_2},t),\hat{\cal R}(\boldsymbol{x},t)\right]\theta^i_{3\boldsymbol{x}}\left [ \hat{\cal R}(\boldsymbol{x_3},t),\hat{\cal R}(\boldsymbol{x},t)\right] \rangle+\theta\text{-perms}\,,
\end{aligned}
\end{equation}
which has been adapted to our notation while decomposing $f_i$ into a product of three differential operators $\theta_i$ acting on $\hat{\cal R}$, i.e.\ reflecting various types of interactions such as those in Table \ref{tab:interactions}. This makes apparent the high dependence of the quantum part on products of commutators and so on the $G$ functions. It also emphasises the key contribution of cubic operators $\theta^i_j$ in specifying the time at which these integrals can be neglected. As we have just seen, because partial time derivatives are important and can hamper classicalisation close to horizon.

\subsection{The importance of real space and the bispectrum shape}
\begin{figure}
    \centering    \includegraphics[width=0.85\textwidth]{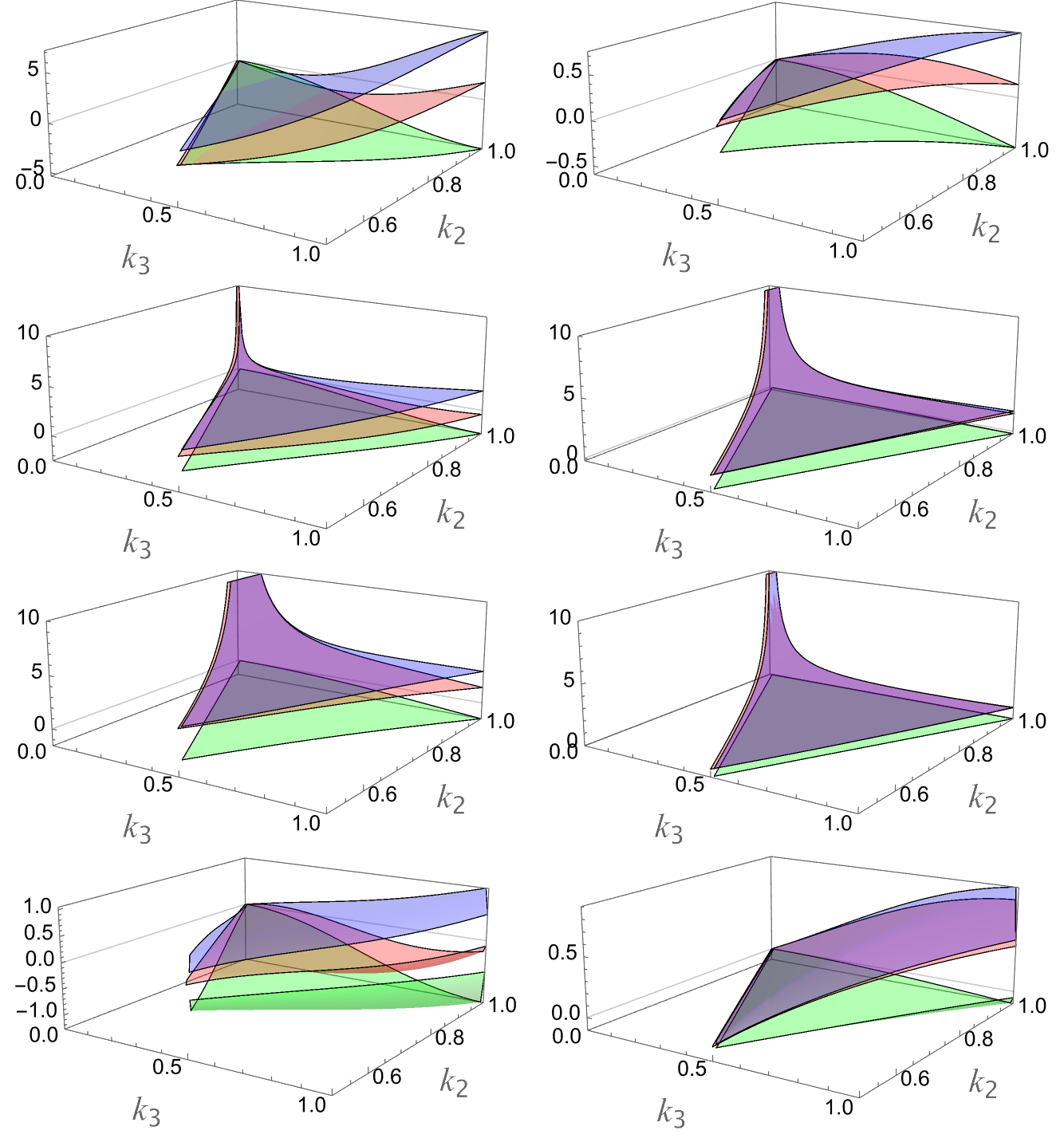}
    \caption{Total (red), classical (blue) and quantum (green) contributions to the shape function of ${\cal C}_{late}$ for ${\cal N}_0 = -1.0$ (left, equivalent to $\eta_0 \simeq -2.72 $) or ${\cal N}_0 = -0.25$ (right, equivalent to $\eta_0 \simeq -1.28 $), to ${\cal N}_f = 4.6$ (equivalent to $\eta_f \simeq -0.01 $) for $g_1 a^3\dot{{\cal R}}^3$, $g_2 a^3{\cal R} \dot{{\cal R}}^2$, $g_3 a{\cal R}(\partial{\cal R})^2$, and $g_4a^3\dot{\cal R}(\partial{\cal R})(\partial\partial^{-2}\dot{\cal R})$ in quasi-dS (top to bottom). $k_1$ is set to $1$ and is the last mode to cross the horizon ($k_{2,3}\leq k_1$), time at which we set the time origin ($k_1=aH$ at $\eta = -1$ or ${\cal N} = 0$  equivalently). Plotting threshold is set at an absolute value of $10$.}
    \label{fig:3Dshape}
\end{figure}

The previous diagnostics have all been defined in Fourier space. However, when returning to real space, the different three-point correlators computed previously will be affected by different tetrapyd regions in Fourier space. In particular, squeezed, folded and equilateral shapes will weigh in with distinct contributions. The shape function, $e_3^2{\cal B}$, is the actual amplitude of the Fourier transform's integrand, with two late-time contributions plotted on a tetrapyd slice in Figure \ref{fig:3Dshape}, first, well before horizon-crossing for the final mode and, secondly, only just beforehand. The standard gravitational interaction terms from Table 1 (i.e., shapes 2--4) have an almost vanishing quantum contribution as horizon-crossing is approached. The most local shape (the third with 
${\cal R}_{\pm}(\partial{\cal R}_{\pm})^2$) indicates that there is only a small quantum contribution even at ${\cal N}_0 = -1.0$. In fact, this small equilateral component is very subdominant relative to the highly peaked squeezed limit where most of the signal resides, which is essentially classical by this time.  This means that from an empirical perspective, when measuring whether this signal is present in CMB or other data, it would still be a good approximation to take an even earlier sub-Hubble prediction of this local shape.  In practical terms for stochastic inflation, this local signal could thus be introduced earlier than naively expected. In other words, the off-shell parts of the tetrapyd could be neglected earlier, which are erased in the integrand back to real space.  We note that these points may also be relevant for multifield inflation models, many of which are expected to exhibit dominant squeezed contributions. 

These observations are in stark contrast to the most equilateral shape (the first EFT shape with $\dot{\cal R}_{\pm}^3$, top of Figure \ref{fig:3Dshape}), which has substantial quantum contributions both at earlier and later times.  On the other hand, the other interactions (2 and 4) with a second time derivative are intermediate between the two extreme cases, but with both being essentially classical just prior to horizon-crossing.  The qualitative behaviour of the squeezed shape (2) is qualitatively the same as the first squeezed counterpart (1), with the signal primarily classical at the early time. The final equilateral shape (4) is quite distinct from the familiar equilateral (1), indicating that there can be some variation in the quantum evolution of the same shape.

Nevertheless, another pertinent point is that of UV divergences that arise when coming back to real space (such as from the coincident limit). This is due to the vacuum energy, also arising in Minkowski space and which might require a regularisation scheme. In particular, when thinking of simulating stochastic initial conditions in general relativity, one cannot go too sub-Hubble because of the risk of introducing divergent non-perturbative noise. One needs to subtract the vacuum energy and only keep its late products. For the same reason, the previous calculations should not look too far into the sub-Hubble regime at the risk of just investigating  already known oscillating divergences. In a regularisation such as using a cutoff or an adiabatic procedure \cite{Durrer09,Pla24}, these far past contributions would be taken out anyway. For this reason and because of the difficult inverse Fourier integration, one should probably be cautious about jumping to conclusions that state classicality in real space earlier than in Fourier space.

\subsection{Diagnosis of simulations and stochastic inflation \label{sec:SI}}

Let us assume that we can generate initial conditions for a patch of universe using linear theory, i.e.\ following a 2-point function $F^s$ at a given initial time for all wavenumbers we are interested in within a box of finite size. The simulation will then take care of evolving $F^s$ further in time and generate interactions terms, all of this on-shell as no quantum simulation exists yet. If the perturbative regime remains, then the previous ${\cal C}^{cl}_{late}$ contributions will be the ones found up to numerical accuracy and up to next order in perturbation theory if $\eta_0$ is the starting time of the simulation. One should of course not forget that such a simulation will not actually separate orders in perturbation theory.

In this framework, the contributions which happened prior to the initial time and are now propagating to the final surface (or those reinteracting for higher orders in perturbation theory) are missing, and obviously any quantum interaction, early or late also.
For the former, it does not matter because our box is finite anyway and can be chosen so that none of these scales are actually within the spectral range of the simulation box. Only second order interactions would be missed.
In theory, one could take into account the early contributions by giving next-order and non-Gaussian ICs using the analytical $n$-point correlators and observe these second order interactions. In practice, introducing these beyond two points with random draws is challenging, though feasible.

The real issue with providing initial conditions for the box's full range of frequencies is the loss of control in the entry time of each mode. In the same way as the previous integrals were computed for some $\eta_0$ chosen for given momenta, the entry of a mode in a simulation must be relative to the mode. If not doing so, each mode is not treated equally in its evolution, thus breaking quasi-scale invariance. This is why stochastic inflation is well-motivated and necessary.

Stochastic Inflation (SI) was introduced in  \cite{Starobinsky_stochastic_1988}, and used as a non-perturbative methodology for light scalars in de Sitter in \cite{starobinsky_equilibrium_1994}.
In stochastic inflation, each  mode that passes a classicality criterion enters the IR system as a stochastic kick \cite{Starobinsky_stochastic_1988, salopek_stochastic_1991,pinol_manifestly_2021, Launay24}. This criterion is usually for the modes to be super-Hubble. Linking this formalism with the previous diagrammatics is possible as evoked in \cite{vanderMeulen2007, Garbrecht:2013coa, Garbrecht:2014dca}.

The formalism has known various treatments with different underlying assumptions. It is however likely that the most rigorous approach extracts the IR sector using a Wilsonian effective field theory, integrating out the UV and the UV-IR mixings to yield an IR action and its path integral. Performing that operation in our framework would mostly entail the appearance of an IR quadratic term $({\cal Q}_q^>)^2$ , schematically
\begin{equation}
    Z \propto \int {\cal D Q}_{cl}^>{\cal D Q}_{q}^>\exp{\left (\displaystyle -i\int d^4 x {\cal Q}_{q}^>\frac{\delta S}{\delta {\cal Q}}[{\cal Q}_{cl}^>] -\int d^4 x d^4 x' {\cal Q}_q^>(x)D(x,x'){\cal Q}_q^>(x')\right )}
    \label{eq:generalSI}
\end{equation}
where ${\cal Q}$ is a non-perturbative gauge-invariant quantity\footnote{Examples of such quantities can be found in \cite{rigopoulos_separate_2003,langlois_evolution_2005} and are vector fields instead. Note that one would need to deal with removing gauge redundancies in the path integral but we chose to stay simple.} linearising as ${\cal R}$, or a non-perturbative scalar in a given gauge as in \cite{pinol_manifestly_2021}.\footnote{This partition function is fully non-perturbative.} The amplitude $D(x)$ comes from the UV dynamics and the coarse-graining procedure, for which perturbation theory can usually be used. Without any Wilsonian UV integration or specific initial conditions, there is to our knowledge no way to get such vertex because of the asymmetry in a raw Keldysh basis change such as in Section \ref{sec:calc}. %Another way to see it is to follow intuitions from Minkowski space where it is believed that the  $({\cal Q}_q^>)^2$ is the only relevant or marginal operator in quasi de Sitter \cite{green_eft_2022}.

The derivation of eq. \eqref{eq:generalSI} has been performed in \cite{morikawa_dissipation_1990,moss_effective_2017,pinol_manifestly_2021} with different notations and typically implies the transfer of initial conditions as implicit boundary conditions to their propagated linear contributions once classical, which are the super-Hubble ones. Eq. (\ref{eq:generalSI}) can be written via the Hubbard-Stratonovich trick \cite{Stratonovich57,Hubbard59} as
\begin{equation}
\resizebox{\textwidth}{!}{%
$\begin{aligned}
     Z & \propto \int{\cal D }\xi\exp{\left(-\frac{1}{2} \int d^4 x d^4 x' \xi(x) D^{-1}(x,x')\xi(x')\right)}\int {\cal D Q}_{cl}^>{\cal D Q}_{q}^>\exp{\left (\displaystyle -i\int d^4 x {\cal Q}_{q}^>(\frac{\delta S}{\delta {\cal Q}}[{\cal Q}_{cl}^>]-\xi)\right )},\\
     & \propto \int{\cal D }\xi\exp{\left(-\frac{1}{2} \int d^4 x d^4 x' \xi(x) D^{-1}(x,x')\xi(x')\right)}\int {\cal D Q}_{cl}^>\delta\left[\frac{\delta S}{\delta {\cal Q}}[{\cal Q}_{cl}^>]-\xi\right],
\end{aligned}$
}
    \label{eq:HStrick}
\end{equation}
which leads to a similar stochastic equation as for the CSA but with contributions at all times. This feature is visible because of the non-zero  stochastic kernel of the free theory $F^{SI}(x_1,x_2) = \langle {\cal Q}_{cl}^{>,(0)}(x_1){\cal Q}_{cl}^{>,(0)}(x_2)\rangle$.
It can be shown \cite{morikawa_dissipation_1990,moss_effective_2017, pinol_manifestly_2021} that $D$ relates to the full quantum free theory kernel 
\begin{equation}
    D(k,t,t^\prime) = {\cal D}_{k,t}{\cal D}_{k,t^\prime}\omega_k(t)\omega_k(t^\prime)F_{\cal Q}(k,t,t^\prime)\,,% = \langle \omega_k(t)\hat{\cal R}_k(t)\omega_k(t')\hat{\cal R}_k(t')\rangle
    \label{eq:eqDSI}
\end{equation}
where $\omega_k$ is a time-dependent window function selecting only modes smaller than a certain $k_\sigma(t)=\sigma a(t)H(t)$, ${\cal D}$ is the free operator, and $F_{\cal Q}(k,t,t^\prime) = \langle {\cal Q}_{cl}^{(0)}(\boldsymbol{k},t){\cal Q}_{cl}^{(0)}(-\boldsymbol{k},t^\prime)\rangle$. Focusing now on a perturbative framework for the IR too, $F^{SI}$ can be calculated as
\begin{equation}
    F^{SI}(x,y)=\int d^4x_2d^4x_3d^4y_2d^4x_3 G(x_2,x)G(y_2,y){\cal D}_{x_2}{\cal D}_{y_2}W(x_2,x_3)*F_{\cal Q}(x_3,y_3)*W(y_2,y_3),
    \label{eq:eqFSIdef}
\end{equation}
using the convolution operator $*$ in real space. By using integration by parts and the definition of the Green's function, this implies that $F^{SI}$ is related to $F$ through
\begin{equation}
    F^{SI}(k,t,t') = \omega_k(t)\omega_k(t')F_{\cal Q}(k,t,t').
    \label{eq:eqFSI}
\end{equation}
This is basically all one needs to build a theory of stochastic inflation. From that, we can calculate the bispectrum using the Keldysh stochastic rules introduced previously but using this new $F^{SI}$. In particular, in the case where the window is a Heaviside function $\omega_k(t)= \Theta(k_\sigma-k)$, it is completely equivalent to computing late contributions ${\cal C}_{late}^{cl}$ as in section \ref{sec:finite} where $\eta_0$ is the time at which the mode with greatest $k$ is equal to $k_\sigma$, which is around horizon crossing. In our language, stochastic inflation is the on-shell trajectory of fluctuations which have crossed the horizon.

Consequently, this also means that the previous ratios tell us how accurate stochastic inflation is with respect to the bispectrum by assuming its classicality, as long as the scales still behave perturbatively. The further away from horizon crossing the better for classicality, but, as we just explained, it doesn’t mean that being close to the horizon is not under control. In particular, key interactions happen around horizon crossing and are usually neglected in the usual stochastic treatments which require super-Hubble cutoffs to invoke the separate universe approximation. Furthermore, it was recently pointed out that associated effects, notably non-negligible gradients, can be crucial beyond single phase slow-roll inflation \cite{Figueroa22,cruces_stochastic_2022,jackson2023,Launay24}. At least in a slow-roll phase, the above computations have shown that classicality is not the reason one should invoke to justify a super-Hubble cutoff, but rather a will for analytical simplicity.

We have made explicit which part of perturbation theory stochastic inflation can reproduce. In particular, as found in \cite{vennin_correlation_2015}, the effectively very classical squeezed limit of \cite{Maldacena_2003} is indeed reproduced by stochastic inflation in a perturbative scenario. 

Despite requiring a perturbative entry of UV modes, stochastic inflation is mostly used to probe non-perturbative IR scenarios with no need of quantum gravity (such as ultra slow-roll, secular growth and non-perturbative IR couplings). In that case, there is even less point in waiting for the decay of quantum interactions to let modes enter the IR system: non-perturbative growth will likely enhance the dominating linear theory and consequently non-perturbative $n$-spectra will follow and dominate any non-gaussian initial conditions with very little influence from perturbative interactions which happened before crossing. Note that this seems to only have been studied in de Sitter space \cite{cohen_stochastic_2021}.

\section{Conclusion}\label{sec:Conclusion}

In this work we examined when the bispectrum can be derived from classical equations of motion as a proxy for classicality. As the review section tried to highlight, there is a rich but incomplete literature on the subject. In particular, confusion arises quickly when it comes to defining what ‘quantum’ means in cosmology. Following earlier work \cite{vanderMeulen2007, Garbrecht:2013coa, Garbrecht:2014dca, Salcedo24}, we applied the field theory tools built by condensed matter and statistical physics communities to the case of an inflating universe with fluctuations that originated from a quantum vacuum.

In doing so, we clarified the identification of the classical (on-shell) and perturbative evolution of this noise through stochastic PDEs and their diagrammatics. For the convenience of cosmologists, we also adapted the usual \textit{in-in} techniques for this matter. In particular, we derived analytically the classical and purely quantum full slow-roll inflating GR contributions to the three-point correlation function in Fourier space for the first time. These expressions allowed us to imagine what a classical evolution would look like for the bispectrum if performed from the far past, noting the important insight that the consistency relations seem to be due to the on-shell evolution only.

A purely classical evolution of a Bunch-Davies vacuum from the far past is of course unrealistic if the universe is as quantum as QFTCS claims. However, truncating this formalism allowed us to focus on times where the quantum contributions are known to decay and where it becomes realistic to neglect them. Comparing the Keldysh classical and non-classical interactions in what we defined as quantum interactivity, we managed to show that common interactions are safe to work with classically even before horizon crossing, while more exotic terms, such as those arising from non-canonical sound speeds, can be treated classically slightly after horizon crossing. This is the first quantitative study of its kind, as past works focused on general arguments to justify that classicality was roughly and safely reached at late times, rather than the detailed nature of this process. The bispectrum is the lowest order interactive statistic to compute but these conclusions can surely be extended to higher $n$-point correlation functions, despite being at higher orders in the coupling’s perturbation theory. Similarly, going beyond the cubic interaction term in GR is also higher order in perturbation theory. Focusing here on the bispectrum is thus a good leading order probe of the quantum interactivity and leads to quantitative results beyond free theory and without requiring loop calculations.

To tackle questions which motivated the need for this study, we have provided insights for numerical purposes and for empirical measurements. If one wants to simulate an inflating patch of universe and look at correlation functions, it is clear that the analytical comparison at first order lies in the truncation of the \textit{in-in} integrals. This is valid for both evolutions of either an initial range of modes or the equivalent open systems of these modes crossing the horizon sequentially. The latter, called stochastic inflation, is mostly used for investigating the non-perturbative growth of perturbations. Our work mostly focussed on early classicality during perturbative phases, thus negating the need to go to super-Hubble where the separate universe approximation ultimately struggles to be accurate anyway.

Finally, this work comes up with clarified methods and new calculations which give a better understanding of statistical signals. The shapes from classical evolution of inflation are now known and can be calculated for many other models. We hope to do so in the near future using the Keldysh decomposition in numerical estimates of the in-in formalism \cite{clarke_probing_2021,LRSZ}.

\acknowledgments
We would like to thank Thomas Colas, Ciaran McCulloch and Bowei Zhang for their expertise and helpful discussions. Gratitude is also extended to everyone whose thoughtful questions sparked the idea for this small project and helped guide it forward.

 Y.L. is supported by the STFC DiS-CDT scheme and the Kavli Institute for Cosmology, Cambridge. E.P.S.S. acknowledges funding from STFC Consolidated Grant No. ST/P000673/1.
 
\appendix

\section{CTP formalism \label{app:CTP}}

The closed-time path formalism allows to relate expectation values like eq.\ \eqref{eq:ininqcl} to the path integral generating functional of the theory \cite{Weinberg95,Chen17diag} 
\begin{eqnarray}\label{eq:in-in_Z0}
Z[J_+,J_-] &=&  \int {\cal D} {\cal R}^{\rm in}_+ {\cal D} {\cal R}^{\rm in}_-\,\rho_0[{\cal R}^{\rm in}_+,{\cal R}^{\rm in}_-] \nonumber \\
 \!\!\!\!\times && \!\!\!\!\!\!\!\!\!\!\!\int\limits_{{\cal R}_+^{\rm in},\,{\cal R}_-^{\rm in}} \!\!{\cal D} {\cal R}_+ {\cal D} {\cal R}_-e^{\frac{i}{\gamma} \left[{\rm S}[{\cal R}_+]-{\rm S}[{\cal R}_-]+\int d^4x J_+(x){\cal R}_+(x) - \int d^4x J_-(x){\cal R}_-(x)\right]}\delta[{\cal R}_+^{\rm f}-{\cal R}_-^{\rm f}]   
\end{eqnarray}
normalised as $Z[J,J] = 1$\footnote{Unlike the often unspecified and ignored normalization of the in-out generating functional used for scattering amplitudes.} and where $\rho_0$ is the initial condition density matrix, here that of the vacuum $
    \rho_0[{\cal R}^{\rm in}_+,{\cal R}^{\rm in}_-] = \langle {\cal R}^{\rm in}_+|0\rangle\langle 0 | {\cal R}^{\rm in}_-\rangle
$. Expectation values can be retrieved via  
\begin{eqnarray}
   \langle \bar{T}\hat{\cal R}(y_1)...\hat{\cal R}(y_m)T\hat{\cal R}(x_1)...\hat{\cal R}(x_n) \rangle &=& \nonumber \\
  && \hspace{-4cm}\frac{(i\gamma)^m\delta^m}{\delta J_-(y_1)... \delta J_-(y_m)}\frac{(-i\gamma)^n\delta^n}{i^n\delta J_+(x_1)... \delta J_+(x_n)} Z[J_+,J_-]\Big |_{J_+ =J_-= 0}.
  \label{eq:dZdJ}
\end{eqnarray}
In this expression, one understands the role played by the branches: $+$ fields yield the time-ordered contributions, and $-$ the anti time-ordered ones respectively.\footnote{In fact the path integral only knows about time-ordering and so the two branches are required for expressing real time expectation values.}

\paragraph{Temporal boundaries.} The (non-squeezed) vacuum wave functional takes the Gaussian form of free harmonic states
\begin{equation}
    \langle{\cal R}^{\rm in}_{\pm}|0\rangle = \mathcal{N}\exp\left(-\frac{1}{2}\int d^3\mathbf{x}d^3\mathbf{y}\, {\cal R}^{\rm in}_{\pm}(\mathbf{x})\frac{\mathcal{E}(\mathbf{x},\mathbf{y})}{\gamma}{\cal R}^{\rm in}_{\pm}(\mathbf{y})\right),
    \label{eq:vacwaveapp}
\end{equation}
for some appropriate kernel $\mathcal{E}(\mathbf{x},\mathbf{y})$ where we explicitly displayed the dependence on $\gamma$.
%\footnote{Note that more generally, the functional includes the conjugate momentum space but can be integrated out in this case.} Implictly, we have defined $t_{\rm in}$ such that ${\cal R}_{\pm}(t_{\rm in})={\cal R}^{\rm in}_{\pm}$. This time can be finite or infinite.

At the other end of the two branches, the delta functional imposes the equality of ${\cal R}_+$ and ${\cal R}_-$ at some future final time $t_f$ after all times of interest; it can be left as a free parameter to be set as the largest time of interest or even taken to $+\infty$. This late time $\delta$-functional joining the $\pm$ branches and imposing equality of ${\cal R}_+$ and ${\cal R}_-$ a the final time $t_{\rm f}$ can be represented as
\begin{equation}
    \delta[{\cal R}_+^{\rm f}-{\cal R}_-^{\rm f}] \propto \exp\left[-\int d^3\mathbf{x} \, \frac{\left({\cal R}_+(\mathbf{x},t_{\rm f}) - {\cal R}_-(\mathbf{x},t_{\rm f})\right)^2}{C^2}\right]
\end{equation}
for some infinitesimal $C \rightarrow 0$. 
In most works, only the ${\cal R}_{\pm}$ path integrals are written and these temporal boundaries are confusingly kept implicit in the paths ensemble. The reason for that is that they are only needed for finding the propagators of the theory, which can be found independently from the path integral, while the path integral is used for interactive calculations.

However, these terms can also be included, despite using only one path integral per branch. All it takes is using a Dirac delta function for each temporal boundary
\begin{equation}
\left\{
    \begin{aligned}
Z[0,0] &= \int {\cal D} {\cal R}_+ {\cal D} {\cal R}_-e^{\frac{i}{\gamma} \int_{\rm in}^{\rm f}dt\int d^3\boldsymbol{x}d^3\boldsymbol{y}\left\{{\cal S}[{\cal R}_+(x)]-{\cal S}[{\cal R}_-(x)]+A^{\rm f}[{\cal R}_{\pm},x]\right\} \delta[\boldsymbol{x}-\boldsymbol{y}] +A^{\rm in}[{\cal R}_{\pm},x,y]},\\
 A^{\rm in}[{\cal R}_{\pm},x,y] & = \frac{i}{2} \mathcal{E}(t,\mathbf{x},\mathbf{y})\left [{\cal R}_{+}(x){\cal R}_{+}(y)+{\cal R}_{-}(x){\cal R}_{-}(y)\right ]\delta(t-t_{\rm in}),  \\
 A^{\rm f}[{\cal R}_{\pm},x] & = i\gamma C^{-2}\left[{\cal R}_+(x) - {\cal R}_-(x)\right]^2\delta(t-t_{\rm f}), 
\end{aligned} \right .
    \label{eq:in-in_Z}
\end{equation}
Under this form, the integral has no constraint on initial and final values as these have been transferred to the integrand. Note that $t_{\rm in}$ and $t_{\rm f}$ need to be finite for the Dirac delta functions to be defined. However this can be extended using the identity commonly referred to as $\epsilon$ prescription \cite{Weinberg95, Kaya19}
\begin{equation}
 \lim_{\substack{\epsilon \rightarrow0}} \int^{u_f}_{-\infty}\epsilon e^{\epsilon u}f(u)du = f(-\infty)\,.
 \end{equation}
This means that, for instance, the previous $\delta(t-t_{\rm in})$ can be substituted by $\epsilon e^{\epsilon t }$ and, when all time integrals are performed in any calculation such as in \eqref{eq:dZdJ}, the limit $\epsilon \rightarrow 0$ can be taken. This can also be done for the future boundary and is very common in scattering studies where particles come from both past and future infinite boundaries \cite{Weinberg95}. For now, we will keep the finite time expression for simplicity and this will not alter any of our generalities.

%Note that field correlators of this sort include both n-point functions as well as Green functions when involving fields that lie on both the $+$ and $-$ branches. The initial and boundary conditions included in the above generating functional lead to the "joining" of the two evolution branches and the characteristic $2\times 2$ matrix structure of the in-in or Schwinger-Keldysh propagators. In our case the initial conditions are those of the Bunch-Davies vacuum. 

\paragraph{Propagators.} The propagators of the associated free theory in this $\pm$ basis of fields are usually denoted $G^{++}, G^{--}, G^{+-}$ and $G^{-+}$ and can be found in the literature, for instance in the cosmological case \cite{Fulling89, Chen17diag}. The propagators are obtained from the free theory action which can in general be written as
\begin{equation}
    {\cal S}[{\cal R}_+(x)]-{\cal S}[{\cal R}_-(x)] = \text{boundary terms} + \frac{1}{2}{\cal R}_+{\cal D}_0{\cal R}_+ -\frac{1}{2}{\cal R}_-{\cal D}_0{\cal R}_-,
\end{equation}
in terms of the free differential operator ${\cal D}_0$ and boundary terms arising from temporal integrations by parts.\footnote{In general, it is assumed that \emph{spatial} boundary terms do not contribute, either for topological reasons, e.g.~implying periodic boundary conditions, or because all dynamical fields are assumed to decay sufficiently fast at spatial infinity.} Such temporal boundary terms are usually ignored or hidden in the literature; the propagators are defined via a $2\times 2$ matrix which is the functional and matrix inverse
\begin{equation}
\left(\begin{array}{cc}\mathcal{D}_0& 0\\ 0&-\mathcal{D}_0\end{array}\right)_{\!\!(x)}
   \left(\begin{array}{cc}{ G^{++}}(x;x^\prime)&{ G^{+-}}(x;x^\prime)
              \\
          {G^{-+}}(x;x^\prime) & {G^{--}(x;x^\prime) }\end{array}\right)
    =\begin{pmatrix} \, \delta^4(x-x^\prime) & 0 \\0 &  \, \delta^4(x-x^\prime)
    \end{pmatrix}\,,
    \label{eq:green+-}
\end{equation}
and are identified with appropriate 2-point functions of $\hat{\cal R}$
\begin{eqnarray}
iG^{++}(x;x^\prime\,)
  &=& \langle 0| T \hat{\cal {\cal R}}(x)\hat{\cal {\cal R}}(x^\prime)| 0 \rangle
      = \theta(x^0\!-\!x^{\prime 0})iG^{-+}(x;x^\prime)
        + \theta(x^{\prime 0}\!-\!x^0)iG^{+-}(x;x^\prime)
\label{Feynman prop}
\\
iG^{--}(x;x^\prime\,)
  &=& \langle 0|\bar T \hat{\cal {\cal R}}(x)\hat{\cal {\cal R}}(x^\prime)| 0\rangle
      = \theta(x^0\!-\!x^{\prime 0})iG^{+-}(x;x^\prime)
        + \theta(x^{\prime 0}\!-\!x^0)iG^{-+}(x;x^\prime)
\,,
\label{antiFeynman prop}
\end{eqnarray}
where the Wightman functions $iG^{+-}$ and $iG^{-+}$ are
\begin{equation}
iG^{+-}(x;x^\prime)=\langle 0 |\hat{\cal {\cal R}}(x^\prime) \hat{\cal {\cal R}} (x)| 0 \rangle
\,,\quad
iG^{-+}(x;x^\prime)=\langle 0|\hat{\cal {\cal R}}(x)\hat{\cal {\cal R}}(x^\prime)| 0 \rangle
\,,
\label{Wightman functions}
\end{equation}
and $|0\rangle$ denotes the physical vacuum state. Note that not all of these in-in propagators are independent since, identically
\begin{equation}
    G^{++}+G^{--}-G^{-+}-G^{+-}=0\,.
    \label{eq:sumG}
\end{equation}

The diagonal elements of the propagator matrix in \eqref{eq:green+-} are the time-ordered (Feynman) and anti-time ordered Green's functions of the operator $\mathcal{D}_0$ the first of which appears in usual S-matrix calculations in QFT, the second simply being the Hermitian conjugate of the first. The real difference of the \emph{in-in} formalism compared to more commonly encountered \emph{in-out} formalism used in computations of scattering amplitudes are the off-diagonal elements in \eqref{eq:green+-} which are solutions of the equation of motion:
$\mathcal{D}_0 \, G^{-+}=\mathcal{D}_0 \, G^{+-}=0$. The expressions \eqref{Feynman prop}, \eqref{antiFeynman prop} and \eqref{Wightman functions} do indeed provide a solution to \eqref{eq:green+-}, but the necessity of the off-diagonal terms is not clear from this equation. In fact, solving the system \eqref{eq:green+-} requires the temporal boundary conditions coming from $\rho_0$ and the branch closure ($A^{\rm in}$ and $A^{\rm f}$). We include these temporal boundary conditions below as part of the operator to be inverted, assuming that they are set at finite times, and show that they indeed lead to \eqref{Feynman prop}, \eqref{antiFeynman prop} and \eqref{Wightman functions}. We do this using the Keldysh fields \eqref{eq:redefKeldysh} which make the physical meaning of the initial conditions more trancparent and allow for a direct link to the classical-statistical approximation. 

\paragraph{Solving for Keldysh propagators. \label{app:solveGreen}}
We assume the action is invariant under translations in 3-space, as is the initial Gaussian density matrix of eq. \eqref{eq:in-in_Z}. A Fourier space treatment is therefore ideal to simplify all quadratic forms. At the free theory level, one is therefore looking at simple harmonic quantum mechanics applied to each mode. 

To rewrite the free action, we will exploit having differential operators with spatial  symmetries $O(t, \boldsymbol{x},\boldsymbol{y}) = O(t, \boldsymbol{x}-\boldsymbol{y})$, so that any quadratic term can be written
\begin{equation}
\begin{aligned}
\int d^3\boldsymbol{x}d^3\boldsymbol{y}\phi(\boldsymbol{x})O(t,\boldsymbol{x},\boldsymbol{y})\phi(\boldsymbol{y}) & = \int_{\boldsymbol{x},\boldsymbol{y}}\int_{\boldsymbol{k}_1,\boldsymbol{k}_2,\boldsymbol{k}_3}O(t,\boldsymbol{k}_3)\phi_{\boldsymbol{k}_1}\phi_{\boldsymbol{k}_2}e^{i\boldsymbol{k}_1\cdot \boldsymbol{x}+i\boldsymbol{k}_1\cdot \boldsymbol{y}+i\boldsymbol{k}_3\cdot (\boldsymbol{x}-\boldsymbol{y})} \\
& = \int_{\boldsymbol{k}_1,\boldsymbol{k}_2,\boldsymbol{k}_3}O(t,\boldsymbol{k}_3)\phi_{\boldsymbol{k}_1}\phi_{\boldsymbol{k}_2}\delta[\boldsymbol{k}_1+\boldsymbol{k}_3]\delta[\boldsymbol{k}_2-\boldsymbol{k}_3] \\
& = \int_{\boldsymbol{k}}O(\boldsymbol{k})\phi_{\boldsymbol{k}}\phi_{-\boldsymbol{k}}.
\end{aligned}
\end{equation}
In particular, a local $O(t, \boldsymbol{x},\boldsymbol{y}) = O(t)\delta(\mathbf{x}-\mathbf{y})\rightarrow O(t)$ in Fourier $\mathbf{k}$-space.

The Keldysh basis redefinitions from eq. \eqref{eq:redefKeldysh} take the partition function from \eqref{eq:in-in_Z} to the form 
\begin{equation}
\begin{aligned}
    Z[0,0] & =  \int {\cal D R}^{\rm in}_{cl}{\cal D R}^{\rm in}_{q} \, \rho_0({\cal R}^{\rm in}_{cl}+\frac{\gamma}{2}{\cal R}^{\rm in}_q,{\cal R}^{\rm in}_{cl}-\frac{\gamma}{2}{\cal R}^{\rm in}_q )\\
    & \times\int_{\rm in}^{\rm f} {\cal D R}_{cl}{\cal D R}_{q}\exp{\left (\int d^4 x\frac{i}{\gamma}S[{\cal R}_{cl}+\frac{\gamma}{2}{\cal R}_{q}]-\frac{i}{\gamma}S[{\cal R}_{cl}-\frac{\gamma}{2}{\cal R}_{q}]\right)} \delta[{\cal R}_q^{\rm f}].
\end{aligned}
    \label{eq:generalcubic}
\end{equation}
and the initial state reads explicitly as
\begin{equation}\label{eq:density-Kelsysh1}
    \rho_0 = \exp\left[-\int d^3xd^3y\left(\frac{1}{\gamma}{\cal R}^{\rm in}_{cl}(\boldsymbol{x})\mathcal{E}(t_0,\boldsymbol{x},\boldsymbol{y}){\cal R}^{\rm in}_{cl}(\boldsymbol{y}) + \frac{\gamma}{4}{\cal R}^{\rm in}_{q}(\boldsymbol{x})\mathcal{E}(t_0,\boldsymbol{x},\boldsymbol{y}){\cal R}^{\rm in}_{q}(\boldsymbol{y}) \right)\right]\nonumber 
\end{equation}
where $\mathcal{E}$ was defined in \eqref{eq:BDamp}. To see how the quadratic operator in the exponent of \eqref{eq:generalcubic} should be inverted, also taking onto account the initial density matrix and final delta functional, we re-write the free partition function ($J_+=J_-=0$) in terms of the Keldysh fields \eqref{eq:redefKeldysh} as
\begin{equation}\label{eq:freetotZ}
         Z_0[0,0]  =  \int {\cal D R}_{cl}{\cal D R}_{q}\exp{\left[\frac{i}{2}\int_{t_{\rm in}}^{t_{\rm f}} dt\int_{ \boldsymbol{k}} 
        \, \begin{pmatrix}
        {\cal R}_{cl}\, , \, 
            {\cal R}_{q}
        \end{pmatrix}_{t,\boldsymbol{k}} {\mathcal{D}}(t, k)\begin{pmatrix}
        {\cal R}_{cl} \\ 
            {\cal R}_{q}
        \end{pmatrix}_{t,-\boldsymbol{k}}\right]}  
\end{equation}
where ${\cal D} =  \bar{\mathcal{D}}_0+ \bar{\mathcal{D}}_b$ is the Fourier space sum of the Fourier free propagator matrix
\begin{equation}
    \bar{\mathcal{D}}_0(t,k)  = \begin{pmatrix}
        0 & \mathcal{D}_0  \\ 
             \mathcal{D}_0 & 0 
        \end{pmatrix},
        \label{eq:freetotZk}
\end{equation}
where $\mathcal{D}_0(t,k)  = -\partial_t\left (\frac{az^2}{c_s^2}\partial_t\cdot\right)-\frac{z^2}{a}k^2$, and the Fourier boundary contributions
\begin{equation}
    \bar{\mathcal{D}}_b(t, \boldsymbol{x},\boldsymbol{y})  = \begin{pmatrix}
        \frac{2i}{\gamma}\mathcal{E}(t,k) \delta[t-t_{\rm in}]& az^2(\delta[t-t_{\rm f}]-\delta[t-t_{\rm in}])\partial_t \\ 
             az^2(\delta[t-t_{\rm f}]-\delta[t-t_{\rm in}])\partial_t& \frac{i\gamma}{2}\mathcal{E}(t,k)\delta[t-t_{\rm in}] +i C\delta[t-t_{\rm f}]
        \end{pmatrix},
        \label{eq:Dbk}
\end{equation}
where $\mathcal{E}(t,k) = z^2(t)\omega_k$. The 2-point correlators of  ${\cal R}$ are obtained by finding the inverse of $\mathcal{D}(t,k)$ 
\begin{equation}
{\cal D}(t,\boldsymbol{k})
   \begin{pmatrix}
       K_k(t,t') & G^R_k(t,t^\prime) \\
       G^A_k(t,t^\prime) & H_k(t,t^\prime)
   \end{pmatrix}
    =\begin{pmatrix}  \, \delta[t-t^\prime] & 0 \\0 &   \, \delta[t-t^\prime]
    \end{pmatrix}.
    \label{inversek}
\end{equation}
The 2-point functions are then given by
\begin{subequations}
    \begin{align}
        \langle {\cal R}_{cl}(t,\mathbf{k}) {\cal R}_{cl}(t',\mathbf{k}^\prime) \rangle = i K_k(t,t')\delta(\mathbf{k}+\mathbf{k}^\prime)\\
        \langle {\cal R}_{cl}(t,\mathbf{k}) {\cal R}_{q}(t',\mathbf{k}^\prime) \rangle = i G^R_k(t,t')\delta(\mathbf{k}+\mathbf{k}^\prime)\\
        \langle {\cal R}_{q}(t,\mathbf{k}) {\cal R}_{cl}(t',\mathbf{k}^\prime) \rangle = i G^A_k(t,t')\delta(\mathbf{k}+\mathbf{k}^\prime)\\
        \langle {\cal R}_{q}(t,\mathbf{k}) {\cal R}_{q}(t',\mathbf{k}^\prime) \rangle = i H_k(t,t')\delta(\mathbf{k}+\mathbf{k}^\prime)
    \end{align}
\end{subequations}

The system \eqref{inversek} involves 12 equations, 4 for each of $t = t_{\rm in}$, $t = t_{\rm f}$ and $t_{\rm in}< t <t_{\rm in}$ portions of the time interval $[t_{\rm in}, t_{\rm f}]$. Starting with the evolution times, \eqref{inversek} yields the propagators' evolution
\begin{subequations}
    \begin{align}
        \mathcal{D}_0(t,k)G^A_k(t,t^\prime) & =  \delta[t-t^\prime], \label{eq:sysa}\\
        \mathcal{D}_0(t,k)H_k(t,t^\prime) & = 0, \label{eq:sysb}\\
        \mathcal{D}_0(t,k)K_k(t,t^\prime) & = 0, \label{eq:sysc}\\
        \mathcal{D}_0(t,k)G^R_k(t,t^\prime) & =  \delta[t-t^\prime].\label{eq:sysd}
    \end{align}
\end{subequations}
\eqref{eq:sysa} and \eqref{eq:sysd} show that $G^A$ and $G^R$ are Green's functions, while $K$ and $H$ are solutions to the e.o.m. At $t=t_{\rm f}$ the boundary terms impose
\begin{subequations}
    \begin{align}
\partial_tG^A_k(t,t^\prime)\displaystyle|_{t = t_{\rm f}} & = 0,  \\
\partial_tH_k(t,t^\prime)\displaystyle|_{t = t_{\rm in}} & = 0, \\
\partial_t K_k(t,t')\displaystyle|_{t=t_{\rm f}} +i C^{-2}G^A_k(t_{\rm f},t^\prime) & = 0, \\
\partial_t G_k^R(t,t')\displaystyle|_{t=t_{\rm f}} +i C^{-2}H_k(t_{\rm f},t^\prime) & = 0. 
    \end{align}
\end{subequations}
Recalling that $C\rightarrow 0$ to represent the delta functional at $t_{\rm f}$ these become
\begin{subequations}
    \begin{align}
\partial_tG^A_k(t,t^\prime)\displaystyle|_{t = t_{\rm f}} & = 0, \label{eq:sysfa} \\
\partial_tH_k(t,t^\prime)\displaystyle|_{t = t_{\rm in}} & = 0, \label{eq:sysfb}\\
G^A_k(t_{\rm f},t^\prime) & = 0, \label{eq:sysfc}\\
H_k(t_{\rm f},t^\prime) & = 0. \label{eq:sysfd}
    \end{align}
\end{subequations}
Furthermore, at $t=t_{\rm in}$ we have 
\begin{subequations}
    \begin{align}
    \partial_tG^A_k(t,t^\prime)\displaystyle|_{t = t_{\rm in}} & = \frac{2i}{\gamma a_{\rm in}}\omega_kK_k(t_{\rm in},t^\prime), \label{eq:sysia}\\
\partial_tH_k(t,t^\prime)\displaystyle|_{t = t_{\rm in}} & = \frac{2i}{\gamma a_{\rm in}}\omega_kG^R_k(t_{\rm in},t^\prime), \label{eq:sysib}\\
\partial_tK_k(t,t^\prime)\displaystyle|_{t = t_{\rm in}} & =\frac{i\gamma}{2a_{\rm in}}\omega_kG^A_k(t_{\rm in},t^\prime), \label{eq:sysic}\\
\partial_tG^R_k(t,t^\prime)\displaystyle|_{t = t_{\rm in}} & = \frac{i\gamma}{2a_{\rm in}}\omega_kH_k(t_{\rm in},t^\prime).\label{eq:sysid}
    \end{align}
\end{subequations}

Equations \eqref{eq:sysa} - \eqref{eq:sysid} completely determine the 2-point functions, as we now show: Since  $\mathcal{D}_0(t,k)\varphi(t)=0$ is a second order O.D.E. in time, one sees from eq. \eqref{eq:sysb}, \eqref{eq:sysfb} and \eqref{eq:sysfd} that $$H_k(t,t') = 0\,.$$  
Looking now at the boundary condition in eq. \eqref{eq:sysfa} and \eqref{eq:sysfc}, we can see that these initial conditions together with ${\cal D}_0(t,k)G_k^A(t,t') = 0$ for the portion $t>t^\prime$, are enough to find $G_k^A(t,t')=0$ for $t>t'$. Hence: 
\begin{equation}
G_k^A(t,t') = \text{Advanced Green's function}\,.
\end{equation}
Similarly, from the other end of the time interval we learn from \eqref{eq:sysib} and \eqref{eq:sysid} (recall that $H_k=0$) that $G_k^R(t,t') = 0$ on the $t<t^\prime$ range and hence
\begin{equation}
G_k^R(t,t') = \text{Retarded Green's function}\,.
\end{equation}
On the other halves of their respective time ranges, these Green's functions do not vanish because of the Dirac delta function at $t=t^\prime$. Integrating \eqref{eq:sysa} and \eqref{eq:sysd} from $t=t'-\epsilon$ to $t=t'+\epsilon$ and taking the limit $\epsilon\rightarrow 0$ we find 
\begin{subequations}
    \begin{align}
\frac{az^2}{c_s^2}\partial_t G^A_k\displaystyle|_{t = t^\prime} &  = 1, \label{eq:sysmida} \\
G^A_k(t^\prime,t^\prime) & = 0, \label{eq:sysmidb}\\
\frac{az^2}{c_s^2}\partial_t G^R_k\displaystyle|_{t = t^\prime} & =  -1, \label{eq:sysmidc} \\
G^R_k(t^\prime,t^\prime) & = 0.\label{eq:sysmidd}
    \end{align}
\end{subequations}
Finding the solution of ${\cal D}_0 G^{R/A}=0$ with these conditions then yields $G^R_k(t,t')$ and $G^A_k(t,t')$ for all values of $(t, t') \in [t_{\rm in}, t_{\rm f}]$. For instance, this leads to eqns. \eqref{eq:defGKGR} and \eqref{eq:FGdS} in a quasi-dS framework.

%\gr{GR:Yoann, why switch to conformal time?  

%...which is better expressed in conformal time $\eta$ 
%\begin{subequations}
%    \begin{align}
%\frac{z}{a}\partial_\eta [zG^A_k]\displaystyle|_{\eta = \eta^\prime} &  = 1, \label{eq:sysmida} \\
%G^A_k(\eta^\prime,\eta^\prime) & = 0, \label{eq:sysmidb}\\
%\frac{z}{a}\partial_\eta[z G^R_k]\displaystyle|_{\eta = \eta^\prime} & =  -1, \label{eq:sysmidc} \\
%G^R_k(\eta^\prime,\eta^\prime) & = 0.\label{eq:sysmidd}
%    \end{align}
%\end{subequations}
%}

\noindent  Note that at no point has the Bunch-Davies vacuum been used so far: the Green's functions encode dynamics only. The Advanced Green's function $G^A(t,t')$ (a propagator from the final temporal point) can then be expressed by including a $\theta[t^\prime-t]$ step function factor, while the Retarded Green's function 
$G^R(t,t')$ (propagating from the initial time) can be expressed by including a $\theta[t-t^\prime]$ step function. We note that $G^R$ and $G^A$ are not independent since $G^R(t,t') = G^A (t',t)$.

Knowledge of the Green's function $G^A$ or $G^R$ along with the initial state (or, more generally, the density matrix) then determines $K_k(t_{\rm in},t')$ and $\partial_tK_k(t,t^\prime)\displaystyle|_{t = t_{\rm in}}$ via 
eqns. \eqref{eq:sysia} and \eqref{eq:sysic} which for our initial state can be written as 
\begin{subequations}
    \begin{align}
    K_k(t_{\rm in},t^\prime) & =-i \frac{\gamma a_{\rm in}}{2\omega_k}\partial_tG^A_k(t,t^\prime)\displaystyle|_{t = t_{\rm in}} \,, \label{eq:K_ini1}\\
\partial_tK_k(t,t^\prime)\displaystyle|_{t = t_{\rm in}} & =i\frac{\gamma \omega_k}{2a_{\rm in}}G^A_k(t_{\rm in},t^\prime)\,. \label{eq:K_ini2} 
    \end{align}
\end{subequations}
These equations provide the initial conditions at $t_{\rm in}$ which determine $K_k(t,t')$ as the appropriate solution of \eqref{eq:sysa} for all values of $(t, t') \in [t_{\rm in}, t_{\rm f}]$. When equations \eqref{eq:K_ini1} and \eqref{eq:K_ini2} are evaluated at $t^\prime = t_{\rm in}$, we can use the previous equations to confirm that $K_k(t_{\rm in},t_{\rm in}) =  -i\frac{\gamma}{2\mathcal{E}(t,k)}$ and $\partial_t K_k(t,t_{\rm in})\displaystyle|_{t=t_{\rm in}} = 0$ as they should for our initial state. Note that, as can be seen from eqns \eqref{eq:sysia} and  \eqref{eq:sysic}, the cl-cl part of the initial density matrix determines the initial amplitude of $K$ while the q-q part determines the initial velocity of $K$. This fact will also be relevant in the formulation of the classical-statistical approximation below.  

The free propagators $G^R$, $G^A$ and $K$ determined above are used in the Feynman diagrams of perturbation theory. We now discuss vertices and the classical-statistical approximation to the dynamics, both linear and non-linear.

\section{Classical statistical approximation \label{app:CSA}}

If one performs an additional integration by parts in eq. \eqref{eq:freetotZ}, the free ${\cal R}_{cl}{\cal D}_0{\cal R}_q$ term can be turned into another ${\cal R}_{q}{\cal D}_0{\cal R}_{cl}$ term, showing that ${\cal R}_{cl}$ is the on-shell part of the field (the one satisfying the equation of motion in the $\gamma \rightarrow 0$ limit). Indeed, in this limit, ${\cal R}_q$ can be integrated out and the exponential becomes a functional Dirac function evaluated at the EOM's left-hand side. This remains valid with interaction terms. This respectively means that ${\cal R}_q$ is the off-shell part of the dynamics\footnote{Not necessarily quantum effects, just any trajectory violating the strict EOM.}. This is why the Keldysh basis is so useful and meaningful.

\paragraph{Classical statistical evolution. }
We have seen that our quasi-dS scenario has a semi-classical approximation only valid for the modes which have crossed the Hubble radius. Let us put that aside for now and see what happens if still performing that approximation. 

In this limit, we take $\gamma \rightarrow 0$ up to second order and so only the leading vertex ${\cal R}_q{\cal R}_{cl}^2$ remains. This all equates to having on-shell evolution only. 
% The second part of the stochastic approximation is about the initial conditions: assuming classicality entails $\mathcal{R}^{\rm in}_{cl} = 0$. 
Overall, the stochastic partition function writes
% classical limit then arises as $\gamma\rightarrow 0$: the "quantum" $\lambda(t){\cal R}_{q}^3$ vertex  becomes suppressed and the density matrix \eqref{eq:density-Kelsysh1} restricts $\mathcal{R}^{\rm in}_{cl} \rightarrow 0$ while the range of the $\mathcal{R}_q^{\rm in}$ integration extends to $(-\infty, +\infty)$. This allows us to write $Z$ as
\begin{equation}
\begin{aligned}
        Z^{sto}[0,0] & =  \int{\cal D R}_{cl}{\cal D R}_{q}\exp\Bigg(i\int d^4 x d^4 y 
    \, {\cal R}_{q}(x)\left(\mathcal{D}_0{\cal R}_{cl}+3\lambda(t){\cal R}_{cl}^2\right)_{y}\delta[x-y]\Bigg) \\
    & \times \exp\Bigg[- \frac{1}{2}\int d^4 x d^4 y  \, \begin{pmatrix}
        {\cal R}_{cl} 
            {\cal R}_{q}
        \end{pmatrix}_{t,\boldsymbol{x}} 
        \begin{pmatrix}
        4{\cal A}_{\rm in}(x,y) & 0 \\
        0 & \gamma^2{\cal A}_{\rm in}(x,y) 
        \end{pmatrix}
        \begin{pmatrix}
        {\cal R}_{cl} \\ 
            {\cal R}_{q}
        \end{pmatrix}_{t,\boldsymbol{y}}\Bigg] ,
\end{aligned}
    \label{eq:Semi-Cl-partitionR}
\end{equation}
where we have kept only the initial conditions for simplicity, and defining 
\begin{equation}
    \mathcal{A}_{\rm in}(x,y) =  \frac{1}{2\gamma}\delta(t_x-t_{\rm in})\delta(t_y-t_{\rm in})\mathcal{E}(t_x,\boldsymbol{x},\boldsymbol{y}).
\end{equation}

The Hubbard-Stratonovich transformation \cite{Hubbard59, Stratonovich57} finally allows us to see that this path integral is stochastic
\begin{equation}
   \exp\Bigg( - \frac{1}{2}\int d^4 x d^4 y 
    \, {\cal R}_{q}(x)\mathcal{A}_{\rm in}(x,y){\cal R}_{q}(y) \Bigg) = \int {\cal D} \xi {\rm P}[\xi|\mathcal{A}_{\rm in}] \exp\Bigg(-i\int d^4 x{\cal R}_{q}\xi\Bigg),
\end{equation}
where ${\rm P}[\cdot|\Sigma]$ is the probability density of a Gaussian field with 2-point function $\Sigma$
\begin{equation}
    {\rm P}[\xi|\Sigma] \propto \exp\Bigg( - \frac{1}{2}\int d^4 x d^4 y 
    \, \xi(x)\Sigma^{-1}(x,y)\xi(y) \Bigg).
    \label{eq:probaxi}
\end{equation}
The path integral can now be calculated and, setting $\gamma = 1$ gives
\begin{equation}
 Z^{sto}[0,0] =  \int {\cal D R}_{cl}{\cal D}\xi \,{\rm P}[{\cal R}_{cl}|4{\cal A}_{\rm in}]{\rm P}[\xi|{\cal A}_{\rm in}]\,\delta[\mathcal{D}_0{\cal R}_{cl}+3\lambda(t){\cal R}_{cl}^2-\xi] .
\end{equation}

The above partition function encodes evolution of the classical equations of motion, enforced by the delta functional, of the original initial conditions encoded in $\rho_0$. In fact, $\xi$ encodes the momentum initial condition and similarly another random variable could be used to include the bulk's initial condition from the other Gaussian on ${\cal R}_{cl}$.  This is the classical statistical approximation to quantum dynamics, also described in \cite{Radovskaya2021} for instance.

\paragraph{Stochastic EOM and correlators. }
% Now let's ignore for now that $G$ is not negligible for all scales and let's assume that the dominating terms are the ones with only one ${\cal R}_q$ term (in our example $\frac{1}{4}\lambda(t)\gamma^3{\cal R}_q^3$ vertex would be neglected).
% This is the lowest order of the (perturbative) semi-classical approximation (after the classical evolution) and corresponds to evolving the classical linear equation of motion sourced by the linear 2-points function
% \begin{equation}
%     \frac{\delta S}{\delta {\cal R}}[\tilde{\cal R}] = \tilde{\xi}
%     \label{eq:langevin}
% \end{equation}
% where $\tilde{\xi}$ is chosen such that the free stochastic kernel $F^{stoch.}$ matches the full theory's $F$ \cite{vanderMeulen2007}.
Another way of writing this stochastic path integral is its associated Langevin equation 
\begin{equation}
    \mathcal{D}_0\tilde{\cal R}+\frac{\delta S^{(3)}}{\delta {\cal R}}[\tilde{\cal R}]=\tilde{\xi}.
    \label{eq:langevinlin}
\end{equation}
where $\tilde{\xi}$ has statistics such that, for free theory, the partition in eq.\eqref{eq:probaxi} has the same two-point function as $\tilde{\cal R}$. This corresponds to considering the lowest order in the coupling, which satisfies $ {\cal D}_0\tilde{\cal R}^{(1)}  = \tilde{\xi}$, and assume that the stochastic two-point function 
\begin{equation}
\begin{aligned}
    K^{stoch}(x, x') & \equiv \langle \tilde{\cal R}^{(1)}(x) \tilde{\cal R}^{(1)}(x') \rangle, \\
    & =\int \int d^4yd^4y' G^R(x,y)G^R(x',y')\langle \tilde{\xi}(y)\tilde{\xi}(y') \rangle,\\
    \end{aligned}
\end{equation}
matches exactly the full quantum theory's $K$ defined earlier.
Note that this $G^R$ is the same as in the full theory with $\gamma = 1$. 

It is possible to check that the diagrammatics of this theory are the same as the in-in theory where the $\gamma^{p\geq 1}$ have vertices have been dropped. This has been achieved in \cite{vanderMeulen2007, Garbrecht:2013coa, Garbrecht:2014dca} and we illustrate it below for the case of the bispectrum.  One can get the next order cubic non-linearity from
\begin{equation}
   {\cal D}_0{\tilde{\cal R}^{(2)} = -\frac{\delta S^{(3)}}{\delta {\cal R}}[\tilde{\cal R}^{(1)}}] = -\epsilon^{ijk}\theta_{i}\left[[\theta_{j}\tilde{\cal R}^{(1)}]\theta_{k}[\tilde{\cal R}^{(1)}]\right]
      \label{eq:langevinlin2}
\end{equation}
where the $\theta_i$ are linear differential operators from the cubic interaction $\theta_1 {\cal R}\theta_2{\cal R}\theta_3 {\cal R}\subset S^{(3)}$. $\epsilon^{ijk}$ is a $\pm 1$ tensor summing symmetries ($\epsilon^{i_1jk}=\epsilon^{i_2jk}$) and depending on how many integrations per parts are required to vary $\theta_i{\cal R}$ in the least action principle (e.g. $\nabla^p$ and $\partial_t^p$ require $p$ of them). This stochastic equation is solved using the previous solution

\begin{equation}
\begin{aligned}
\tilde{\cal R}^{(2)}(x) 
& = -\epsilon^{ijk}\int d^4 y G^R(x,y) \theta_{i}\left[\theta_{j}[\tilde{\cal R}^{(1)}]\theta_{k}[\tilde{\cal R}^{(1)}\right](y) \\
& = -\epsilon^{ijk}\int d^4 y\int d^4 z_1\int d^4 z_2 G^R(x,y) \theta_{i,y}\left[\theta_{j,y}[G^R](y,z_1)\theta_{k,y}[G^R](y,z_2)\right] \tilde{\xi}(z_1)\tilde{\xi}(z_2).
\end{aligned}
\end{equation}
For instance, the 3-point function at lowest order (4th order in $\tilde{\xi}$) is given by

\begin{equation}
\resizebox{\textwidth}{!}{
$\begin{aligned}
     \langle \tilde{\cal R}^{(2)}(x_1) \tilde{\cal R}^{(1)}(x_2) \tilde{\cal R}^{(1)}(x_3) \rangle & = - \epsilon^{ijk}\int d^4 y_1 d^4 y_2 d^4 y_3 d^4 z_1 d^4 z_2 \theta_{i,y}\left [\theta_{j,y}[G^R](y,z_1)\theta_{k,y}[G^R](y,z_2) \right ]\\ & \times G^R(x_1,y_1)G^R(x_2,y_2) G^R(x_3,y_3)  \langle\tilde{\xi}(z_1)\tilde{\xi}(z_2)\tilde{\xi}(y_2)\tilde{\xi}(y_3)\rangle \\
     & = - 2\epsilon^{ijk}\int d^4 y G^R(x_1,y)\theta_{i,y}\left[\theta_{i,y}[F](y,x_2)\theta_{j,y}\theta_{k,y}[F](y,x_3)\right]  \\
     &  = -2\int d^4 y \theta_{1,y}[G^R](x_1,y)\theta_{2,y}[F](y,x_2)\theta_{3,y}[F](y,x_3) + perm.
\end{aligned}$
}
\label{eq:demoSTObi}
\end{equation}
once the Isserlis theorem has been applied and keeping only connected contributions and where the integration by parts have been reversed. Note that the full expression also involves summing over $(x_1, x_2,x_3)$ permutations. This result in real space matches the in-in formalism tree-level formula with $FFG$ vertices only. This will be true at all orders in $\tilde{\xi}$ for any $n$-point correlator \cite{Radovskaya2021,vanderMeulen2007} and can also be generalized to non-cubic vertices with one $G$ or more (if odd) if one wants to look at next orders in $\gamma$.

We hope that by now, our vision of what a 'classical' contribution is: a contribution without 'quantum' features, where 'quantum' is what cannot be generated from the classical equation of motion or any of its stochastic modifications.

This section also gives the opportunity to highlight that a path integral doesn’t have to be quantum. This is justified because Schrödinger-like equations (and so path integrals) can be written for Hilbert spaces of stochastic states \cite{,Bounakis20,rigopoulos_fluctuation-dissipation_2013,vastola_stochastic_2019}. Going from the stochastic EOM to the path integral is known as the MSRJ procedure \cite{MSRformula}.

\section{Redefinition contributions to the leading bispectrum \label{app:appRedef}}
Eq. \ref{eq:eqH3} actually constitutes the action of the redefined field ${\cal R}_n$ from \cite{Maldacena_2003, chen_observational_2007}. The total bispectrum is

\begin{equation}
    \langle \hat{{\cal R}}_{\boldsymbol{k}_1}\hat{{\cal R}}_{\boldsymbol{k}_2}\hat{{\cal R}}_{\boldsymbol{k}_3}\rangle =     \langle \hat{{\cal R}}_{n,\boldsymbol{k}_1}\hat{{\cal R}}_{n,\boldsymbol{k}_2}\hat{{\cal R}}_{n,\boldsymbol{k}_3}\rangle + \langle f(\hat{{\cal R}}_{n,\boldsymbol{k}_1})\hat{{\cal R}}_{n,\boldsymbol{k}_2}\hat{{\cal R}}_{n,\boldsymbol{k}_3}\rangle+perm
\end{equation}
where { $f(\hat{{\cal R}}_{\boldsymbol{k}_1}) =\frac{\varepsilon_2}{4 c_s^2} {\cal R}^2+\frac{1}{c_s^2 H} {\cal R} \dot{\cal R} +\frac{1}{2 a^2 H}\left[(\partial {\cal R})(\partial \chi)-\partial^{-2}\left(\partial_i \partial_j\left(\partial_i {\cal R} \partial_j \chi\right)\right)\right] +...
$ is the redefinition chosen in \cite{chen_observational_2007}.}
Only the first term is usually kept at leading order in slow-roll in the super-Hubble limit where derivatives damp the other terms. However, around horizon crossing, these are perfectly legitimate contributions. The reason why we don't consider them is the fact that it is fully accounted for in a classical framework at least at linear order.

\bibliography{mybiblio}
\bibliographystyle{JHEP}
\end{document}